\documentclass[journal,twocolumn,final]{IEEEtran}

\usepackage[english]{babel}
\usepackage{cuted}
\usepackage{flushend}
\usepackage{amsthm}
\usepackage{amsmath}
\usepackage{amsfonts}
\usepackage{dsfont}
\usepackage{mathrsfs}
\usepackage{pifont}
\usepackage{amssymb}
\usepackage{verbatim}
\usepackage{upgreek}
\usepackage{color}
\usepackage[final]{graphicx}
\usepackage{caption}
\usepackage{algorithmic}
\usepackage{algorithm}
\usepackage{epsfig}
\usepackage{eucal}
\usepackage{cite}
\usepackage{subfigure}
\usepackage{stmaryrd}

\makeatletter
\def\maketag@@@#1{\hbox{\m@th\normalfont\normalsize#1}}
\makeatother

\newcommand{\bbm}{\begin{bmatrix}}
\newcommand{\ebm}{\end{bmatrix}}

\newcommand{\bit}{\begin{itemize}}
\newcommand{\eit}{\end{itemize}}

\newcommand{\ben}{\begin{enumerate}}
\newcommand{\een}{\end{enumerate}}

\newcommand{\bdesc}{\begin{description}}
\newcommand{\edesc}{\end{description}}

\newcommand{\bea}{\begin{array}}
\newcommand{\eea}{\end{array}}

\newcommand{\tr}{\mbox{\rm Tr}\, }

\newcommand{\beqa}{\begin{eqnarray}}
\newcommand{\eeqa}{\end{eqnarray}}

\newcommand{\ds}{\displaystyle}

\newcommand{\Comment}[1]{}

\def\C{{\mathds C}}

\def\cC{\mbox{$\CMcal C$}}

\def\cL{\mbox{$\mathcal L$}}
\def\cN{\mbox{$\CMcal N$}}

\def\E{{\mathbb E}}

\newcommand{\be}{\begin{equation}}
\newcommand{\ee}{\end{equation}}

\newcommand{\bzero}{{\mbox{\boldmath $0$}}}

\newcommand{\boa}{{\mbox{\boldmath $a$}}}

\newcommand{\bn}{{\mbox{\boldmath $n$}}}
\newcommand{\bm}{{\mbox{\boldmath $m$}}}
\newcommand{\bp}{\mbox{\boldmath $p$}}
\newcommand{\bor}{{\mbox{\boldmath $r$}}}

\newcommand{\bu}{{\mbox{\boldmath $u$}}}
\newcommand{\bv}{{\mbox{\boldmath $v$}}}
\newcommand{\bx}{{\mbox{\boldmath $x$}}}

\newcommand{\bz}{{\mbox{\boldmath $z$}}}

\newcommand{\bA}{{\mbox{\boldmath $A$}}}
\newcommand{\bB}{{\mbox{\boldmath $B$}}}
\newcommand{\bC}{{\mbox{\boldmath $C$}}}
\newcommand{\bD}{{\mbox{\boldmath $D$}}}

\newcommand{\bI}{{\mbox{\boldmath $I$}}}

\newcommand{\bM}{{\mbox{\boldmath $M$}}}
\newcommand{\hbM}{{\mbox{$\widehat{\bM}$}}}

\newcommand{\bR}{{\mbox{\boldmath $R$}}}
\newcommand{\bS}{{\mbox{\boldmath $S$}}}

\newcommand{\bV}{{\mbox{\boldmath $V$}}}

\newcommand{\bZ}{{\mbox{\boldmath $Z$}}}

\DeclareMathOperator*{\argmax}{arg\,max}

\newcommand{\bnu}{{\mbox{\boldmath $\nu$}}}

\newcommand{\dmax}{\begin{displaystyle}\max\end{displaystyle}}
\newcommand{\dmin}{\begin{displaystyle}\min\end{displaystyle}}

\newcommand{\testconv}{\mbox{$
\begin{array}{c}
\stackrel{ \stackrel{\textstyle H_{1,0}}{\textstyle >} }{
\stackrel{\textstyle <}{\textstyle H_0} }
\end{array}
$}}

\newcommand{\testest}{\mbox{$
		\begin{array}{c}
		\stackrel{ \stackrel{\textstyle H_{1,\hat{i}}}{\textstyle >} }{
			\stackrel{\textstyle <}{\textstyle H_0} }
		\end{array}
		$}}

\title{Adaptive Radar Detection and Classification Algorithms for Multiple Coherent Signals}

\vspace{0.1cm}

\author{Sudan Han, Linjie Yan, Yuxuan Zhang, Pia Addabbo, \IEEEmembership{Senior Member, IEEE},  \\ Chengpeng Hao,
\IEEEmembership{Senior Member, IEEE}, and Danilo Orlando, \IEEEmembership{Senior Member, IEEE}
\thanks{Sudan Han is with  the  National  Innovation  Institute  of  Defense  Technology, Beijing, China. E-mail: xiaoxiaosu0626@163.com.}
\thanks{Pia Addabbo is with Universit\`a degli studi ``Giustino Fortunato'', Benevento, Italy. E-mail: {\tt 
p.addabbo@unifortunato.eu}.}
\thanks{Linjie Yan, Yuxuan Zhang, and Chengpeng Hao are with Institute of Acoustics, Chinese Academy of Sciences,
Beijing, China. E-mail: {\tt yanlinjie16@163.com; zhangyuxuan@mail.ioa.ac.cn; haochengp@mail.ioa.ac.cn}.}
\thanks{Danilo Orlando is with the Engineering Faculty of Universit\`a degli Studi ``Niccol\`o Cusano'', 
via Don Carlo Gnocchi 3, 00166 Roma, Italy. E-mail: {\tt danilo.orlando@unicusano.it}.}
}

\begin{document}

\maketitle

\begin{abstract}
In this paper, we address the problem of target detection in the presence of coherent (or fully correlated) signals, 
which can be due to multipath propagation effects or electronic attacks by smart jammers. To this end, 
we formulate the problem at hand as a multiple-hypothesis test that, besides the conventional radar 
alternative hypothesis, contains additional hypotheses accounting for the presence of an unknown number of interfering signals. 
In this context and leveraging the classification capabilities of the  Model Order Selection rules, we devise penalized 
likelihood-ratio-based detection architectures that can establish, as a byproduct,  which hypothesis is in force. 
Moreover, we propose a suboptimum procedure to estimate the angles of arrival of multiple coherent signals ensuring 
(at least for the considered parameters) almost the same performance as the exhaustive search. 
Finally, the performance assessment, conducted over simulated data and in comparison with conventional radar detectors, 
highlights that the proposed architectures can provide satisfactory performance in terms of probability of detection and correct classification.

\end{abstract}


\section{Introduction}
\label{Sec:Intro}
In recent years, radar systems have become ubiquitous in real life due to the advances in digital 
architectures and miniaturization technologies. More importantly, the huge amount of computational power 
has paved the way for sophisticated processing algorithms fed by digital samples and leading to new architectures 
where the presence of analog hardware resources devoted to a specific task is very limited \cite{Richards,ScheerMelvin,richards2013principles}. As a consequence, modern radar systems are extremely flexible
and can incorporate different functions without additional hardware components.

In system design, the turning point is represented by the knocking down of high-frequency sampler 
and processing board costs that have allowed the development of fully-digital architectures. 
In this evolving scenario, radar research community continues to devise algorithms with increased 
complexity that take full advantage of the potential provided by digital architectures. The {\em quid pro quo}
of the increased computational power is represented by the enhanced performance as corroborated, 
for instance, by the space and/or time adaptive algorithms developed in the open literature. 
As a matter of fact, focusing on the radar detection task and starting from the seminal works by 
Kelly \cite{kelly1986adaptive,kelly-tr}, a plethora of decision schemes have been proposed by enriching
the design assumptions in order to account for a priori information and/or specific aspects of the 
application/system under consideration \cite{Melvin-2000,kelly1986adaptive,robey1992cfar,gini1,RicciRao,
WLiuRao, Yuri01, BOR-Morgan,6757035,LIU2020107268, LiuSun19, CP00, fogliaPHE_SS, HaoSP_HE,
FOGLIA2017131,hongbinPersymmetric}.

It is also important to underline that the benefits coming from the aforementioned technology advances 
have been also exploited by the Electronic Warfare (EW) systems, which have adapted themselves to the more and more
reliable capabilities of radar systems leading to a more effective class of electronic countermeasures 
referred to as smart jammers \cite{barton2013radar,richards2013principles,5751244,7131043,
schleher1999electronic,Neribook}. For instance, modern noise-like jamming systems are capable
of transmitting narrow-band interfering signals which are concurrent with the 
radar pulses according to the radar pulse repetition interval after having estimated it (EW integrated systems). Moreover, they can use
an intrapulse modulation to generate a noise bandwidth at radio frequency, while maintaining phase 
coherence over a group of successive jamming pulses \cite{Neribook}. Remarkably, this kind of jammers
can generate signals that are coherent (i.e., fully correlated) with the desired signal even though
the former impinge on the radar from different directions \cite{575698,215306}. An analogous situation 
occurs in scenarios where, due to multipath propagation, replicas of the original signal come back
to the receiver with a sufficiently small delay difference \cite{215306,1164583}.

The main drawback caused by the presence of multiple coherent signals is that they can completely destroy 
the performance of the most common high-resolution direction finding algorithms for adaptive array 
systems \cite{VanTrees4,1164649}. As a matter of fact, coherent signals appear as
a single signal impinging on the array of sensors and that, more importantly, arrives 
from a direction which is quite different from that of the sought signals.
As a consequence, high-resolution eigenstructure-based techniques as, for instance, 
MUltiple SIgnal Classification algorithm \cite{VanTrees4}, fail to correctly resolve the 
signals jeopardizing the Angle of Arrival (AoA) estimation \cite{575698}.
Similar remarks also hold for another important approach to AoA estimation, namely the CLEAN algorithm \cite{stoica2005spectral} which 
consists in iterative cancellations of strong signals under the assumption that the spatial covariance matrix results from 
the sum of contributions associated with uncorrelated sources. Remarkably, it can be used for the detection of weak targets embedded
in strong interference signals \cite{CleanBistaticRadar1,CleanBistaticRadar2,CleanBistaticRadar3}.

In order to mitigate the effects of coherent signals, data feeding the direction finding algorithm
can be suitably preprocessed in order to decrease the correlation of the impinging signals.
In this respect, spatial smoothing technique \cite{1164649} combines data obtained from synthetic subarrays 
whose size is lower than that of the original array with the drawback of a reduced angular resolution. 
In the context of subspace-based algorithms, in \cite{97999} a method based upon the weighted subspace
fitting criterion is proposed to jointly detect and estimate the number and the related parameters
of the coherent signals. 
However, this method does not allow for the control
of the probability of false alarm ($P_{fa}$) which is of primary concern in radar. 
Alternative approaches against the coherent signals problem 
can rely on the Maximum Likelihood (ML) estimation \cite{7543} which, however, is computationally 
intensive and requires the knowledge of the number of coherent signals, or the compressed sensing paradigm where
a further stage for estimate fusion is required \cite{slim,ECCMYan}.

Thus, in order to save
computational resources (due to the activation of additional processing stages also in situations where they are not required) 
and to take advantage of the full 
angular resolution of the system, a detection stage capable of 
deciding for the presence of a target and possible coherent signals by estimating, as a byproduct, their number
along with other side information is highly desirable. In fact, when this stage declares the 
presence of noncoherent signals, direction finding algorithms can be applied without losses in resolution.
On the other hand, when coherent signals are present, the rough estimates of the corresponding parameters provided by this stage
can be used to drive ML-based direction finding algorithms to reduce their computational load. Remarkably, such
stage can be viewed as an Electronic Counter-CounterMeasure.

With the above remarks in mind, in this paper we devise a detection architecture accounting for the presence
of coherent signals and that can provide an estimate of their number as well as the AoAs. To this end,
at the design stage, we formulate the problem at hand in terms of a multiple hypothesis test,
comprising the usual null (or interference-only) hypothesis, the conventional signal-plus-interference hypothesis, 
and multiple alternative hypotheses 
that differ in the number of coherent echoes. These signals are assumed to follow Swerling II Target Model, 
which assumes that the Radar Cross Section (RCS) of the target obeys the chi-squared distribution with two degrees of 
freedom \cite{Richards}. Then, assuming an upper bound on the number of impinging signals (that can be dictated by system parameters), 
we conceive likelihood-ratio-based decision rules which exploit suitable penalty terms borrowed 
from the Model Order Selection (MOS) rules \cite{Stoica1,StoicaBabu}. Therefore, the
proposed architectures can provide an estimate of the actual number of coherent signals.
As for the AoA estimation, the angular sector under consideration
is sampled to form a discrete set of angular positions that are used at the design stage.
Besides, since the exhaustive search can become very time demanding for high numbers of signals, 
we design a suboptimum iterative procedure providing satisfactory performance at least for the considered
numerical examples. Finally, the performance assessment is conducted over synthetic data in comparison
with classical detection architectures and highlights that the proposed schemes represent an effective
means to face with the problem of target detection in the presence of coherent signals.

The remainder of the paper is organized as follows. Section \ref{Sec:Problem_Formulation} 
is devoted to problem formulation and the definitions used in the next derivations, 
while the design of the detection and estimation architectures is described in
Section \ref{Sec:Architecture_Designs}. Section \ref{Sec:Performance} shows 
the effectiveness of the proposed strategies through numerical examples on simulated data. 
Finally, Section \ref{Sec:Conclusions} contains concluding remarks and charts a course for future works.
	
\subsection{Notation and Acronyms} 
In the sequel, vectors and matrices are denoted by boldface lower-case and upper-case letters, respectively. 
Symbols $\det(\cdot)$, $\tr(\cdot)$, $(\cdot)^T$, and $(\cdot)^\dag$ denote the determinant, trace, transpose, 
and complex conjugate transpose, respectively. 
If $A$ and $B$ are two generic sets, then $A\times B$ denotes the Cartesian product between $A$ and $B$.
As to the numerical sets, $\C$ is the set of complex numbers, and $\C^{N\times M}$ is the Euclidean 
space of $(N\times M)$-dimensional complex-valued matrices 
(or vectors if $M=1$).  The imaginary unit is indicated by $j$.
The Euclidean norm of a generic vector $\bx$ is denoted by $\|\bx\|$ whereas the modulus of a complex number $x$ is denoted by $|x|$. 
The symbol $\E[\cdot]$ denotes statistical expectation while $\bzero$ and $\bI$ are the null vector/matrix and the identity matrix, respectively,
of proper size. 
Given two events $A$ and $B$, the conditional probability of $A$ given $B$ is denoted by $P(A|B)$.
The acronyms PDF and IID mean Probability Density Function and Independent and Identically Distributed, respectively. 
For a given matrix $\bA$, $\lambda_{\max}\{\bA\}$ denotes the maximum eigenvalue of $\bA$. 
Finally, we write $\bx\sim\cC\cN_N(\bm, \bM)$ if $\bx$ is a complex circular $N$-dimensional normal vector 
with mean $\bm$ and positive definite covariance matrix $\bM$. 

\section{Problem Formulation}
\label{Sec:Problem_Formulation}
Let us assume that a search radar system is equipped with a uniform linear array with $N$ antennas and 
transmits $L$ pulses in the nominal beam position during one scan cycle.
Then, each antenna collects 
$L$ samples from the cell under test. Denote by $\bz_l\in \C^{N\times 1}$, $l=1,\ldots,L$, the vector 
whose entries are the $l$th returns from each antenna 
and by $\bZ_L=[\bz_1,\ldots,\bz_L] \in \C^{N\times L}$ the overall data matrix, 
the classical radar detection problem consists in deciding whether or not $\bZ_L$ contains 
the target of interest, namely, a component whose signature coincides with the nominal steering vector.
%\textcolor{red}{
Before proceeding with the problem formulation, it is important to state here that
the columns of $\bZ_L$ are assumed statistically independent and, hence, temporally noncoherent, whereas each 
column is representative of both spatially correlated clutter and spatially coherent useful signal components (a point better explained below).
%}
%It is important to notice that the system considered here
%performs noncoherent processing and, hence, the resulting steering vectors are spatial only and without 
%any Doppler component.

As customary, we assume that a set of $K$ ($K\geq N$) secondary data, $\bz_k\in \C^{N\times 1}$, $k=1,\ldots,K$, free of target components 
and sharing the same statistical properties of the interference in the cell under test, is available \cite{kelly1986adaptive,robey1992cfar,
BOR-Morgan,GLRT-based}. 
Under the above assumptions, this problem can be formulated in terms of a conventional binary hypothesis test 
whose expression is (see \cite{kelly1986adaptive,robey1992cfar,BOR-Morgan} and references therein)
\begin{equation} \left\{
\begin{aligned}
&H_0:\begin{cases}
\bz_l = \bn_l,\ \ l=1,\ldots,L,
\\
\bz_k = \bn_k, \ \ k=1,\ldots,K,
\end{cases}
\\
&H_1:
\begin{cases}
\bz_l = \alpha_l\bv(\theta_t) + \bn_l, \ \ l=1,\ldots,L,
\\
\bz_k = \bn_k, \ \ k=1,\ldots,K,
\end{cases}
\end{aligned}
\right.
\label{conventionalproblem}
\end{equation} 
where 
\begin{itemize}
\item $\bn_l,\bn_k\sim\cC\cN_N(\bzero,\bM)$, $l=1,\ldots,L$, $k=1,\ldots,K$, are IID with unknown positive definite covariance 
matrix $\bM\in\C^{N\times N}$ representative of the thermal noise plus clutter; $\bM$ is referred to in the following 
as Interference Covariance Matrix (ICM);
\item $\alpha_1,\ldots,\alpha_L\sim\cC\cN_1(0,\sigma_{\alpha}^{2})$ with $\sigma_{\alpha}^{2}>0$ are IID random variables
accounting for both the target response (RCS) and channel effects (Swerling II target model \cite{Richards});
\item $\bv(\theta_t)\in\C^{N\times 1}$ is the nominal (spatial) steering vector whose expression is
$$
\bv(\theta_t)\!\!=\!\![1,\exp\{j\pi \sin(\theta_t)\},\ldots,\exp\{j\pi (N-1)\sin(\theta_t)\}]^T\!\!,
$$ 
where $\theta_t$ is the (known) AoA of the target measured with respect to the array broadside and
we have assumed that the inter-element spacing is half the operating wavelength in order to avoid 
the aliasing of the spatial frequency.
\end{itemize}

However, as stated in Section \ref{Sec:Intro}, in practical applications the generic $\bz_l$ might not only contain the direct 
echo but also returns from other directions and that, more importantly, are coherent (i.e., fully correlated) with the former 
due, for instance, to the effects
of multipath propagation and/or the presence of smart jammers \cite{575698,215306,1164583}. 
As a consequence, the conventional radar detection problem \eqref{conventionalproblem} consisting of two hypotheses
may be no longer representative of the actual operating scenario. In fact, in such case, the alternative hypothesis can be 
replaced by the following
\be
\!\!\!H_{1,M}: \!\!\! \ 
\begin{cases}
\ds \bz_l = \alpha_l\bv(\theta_t) + \sum_{i=1}^{M}\beta_{l,i}\bv(\theta_i) + \bn_l, &\!\!\!\!\! l=1,\ldots,L,
\\
\bz_k = \bn_k, & \!\!\!\!\! k=1,\ldots,K,
\end{cases}
\ee
where $M$ is the actual number of coherent signals, which is unknown. 
However, notice that the uncertainty on $M$ naturally leads to multiple alternative 
hypotheses, $H_{1,i}$ say, which differ in the number of coherent signals. Therefore, assuming an upper bound, $M_c$ say, on the latter 
unknown parameter, a possible approach to account for the presence of possible fully-correlated signals consists 
in considering a multiple hypothesis test where 
besides the conventional $H_0$ and $H_1$ hypotheses, $M_c<N$ additional hypotheses\footnote{The constraint $M_c<N$ is due to both the fact
that $N$ degrees of freedom are required to estimate $N$ AoAs also making matrix $\bV_i$, defined in \eqref{eqn:pdf_h1}, full-column rank.},
corresponding to scenarios containing a different number of coherent signals, appear.
Therefore, problem \eqref{conventionalproblem} can be generalized as follows
\begin{equation} \left\{
\begin{aligned}
H_0 &:\begin{cases}
\bz_l = \bn_l, & l=1,\ldots,L,
\\
\bz_k = \bn_k, & k=1,\ldots,K,
\end{cases}
\\
H_{1,0} &:
\begin{cases}
\bz_l = \alpha_l\bv(\theta_t) + \bn_l, & l=1,\ldots,L,
\\
\bz_k = \bn_k, & k=1,\ldots,K,
\end{cases}
\\
H_{1,i} &:
\begin{cases}
\ds \bz_l = \alpha_l\bv(\theta_t) + \sum_{k=1}^{i}\beta_{l,k}\bv(\theta_k) + \bn_l, &\!\!\!\!\!  l=1,\ldots,L,
\\
\bz_k = \bn_k, &\!\!\!\!\!  k=1,\ldots,K,
\end{cases}
\end{aligned}
\right.
\label{problem1}
\end{equation} 
$i=1,\ldots,M_c$, where the $\alpha_l$s, the $\bn_l$s, and the $\bn_k$s have been already defined 
for problem \eqref{conventionalproblem} and
\begin{itemize}
\item $M_c\geq M$ is an upper bound on the number of coherent 
signals\footnote{Note that problem (3) reduces to (1) when $M_c=0$.};
\item given $i\in\{1,\ldots,M_c\}$ and $k\in\{1,\ldots,i\}$, the coefficients $\beta_{1,k},\ldots,\beta_{L,k}\in\C$ 
are IID complex Gaussian random variables with zero mean and 
variance $\sigma_{k}^{2}>0$; they represent the complex amplitudes of the $k$th coherent 
signal over the time\footnote{It is worth noticing that Swerling II target model is also adopted for the coherent signals.};
\item given $i\in\{1,\ldots,M_c\}$, $\theta_1,\ldots,\theta_i$ are the unknown AoAs of the additional coherent signals.
\end{itemize}
Finally, we assume that, $\forall i=1,\ldots,M_c$, the random variables $\alpha_l$, $\beta_{l,1},\ldots,\beta_{l,i}$ are fully correlated. 
Otherwise stated, given $l\in\{1,\ldots,L\}$ and $i\in\{1,\ldots,M_c\}$, 
$\forall k,h\in\{1,\ldots,i\}$ and $k\neq h$ the following equalities hold
\be
\E[\alpha_l\beta_{l,k}]=\sigma_{\alpha}\sigma_{{k}} \quad \mbox{and} \quad \E[\beta_{l,k}\beta_{l,h}]=\sigma_{{k}}\sigma_{{h}}.
\ee
In order to unburden the notation, hereafter we omit the dependence of the steering vectors on the AoAs and 
denote by $\bv_t$ the nominal steering vector whereas we use $\bp_i$ to indicate the steering vector
associated with the coherent signal whose AoA is $\theta_i$.

Before concluding this section, we provide some definitions that will come in handy for the next developments. More precisely,
the PDF of $\bZ_L$ under $H_0$ has the following expression
\be
f_0(\bZ_L;\bM)=\displaystyle{\left[\frac{1}{\pi^N\det(\bM)}\right]^L\exp\left[-\tr\left(\bM^{-1}\bZ_L\bZ_L^
\dag\right)\right]},
\ee
whereas its PDF under the generic $H_{1,i}$, $i=0,\ldots,M_c$, is given by
\begin{multline}
f_{1,i}(\bZ_L;\bM, \bR_i, \bV_i)=\left[\frac{1}{\pi^N\det(\bM+\bV_i\bR_i\bV_i^\dag)}\right]^L\\
\times \exp\left\{-\tr\left[\left(\bM+\bV_i\bR_i\bV_i^\dag\right)^{-1}\bZ_L\bZ_L^\dag\right]\right\},
\label{eqn:pdf_h1}
\end{multline}
where $\bV_i=[\bv_t,\bp_1,\ldots,\bp_{i}] \in \C^{N\times(i+1)}$ and $\bR_i=\E[\bnu_{l,i}\bnu_{l,i}^\dag]\in\C^{(i+1)\times(i+1)}$, 
$l=1,\ldots,L$, with $\bnu_{l,i}=[\alpha_l,\beta_{l,1},\ldots,\beta_{l,i}]^T \in \C^{(i+1)\times 1}$.
It is understood that when $i=0$, $\bV_0$ and $\bnu_{l,0}$ coincide with $\bv_t$ and $\alpha_l$, respectively.

\section{Detection Architecture Designs}
\label{Sec:Architecture_Designs}
In this section, we devise a detection architecture for problem \eqref{problem1} that relies on a {\em penalized} log-likelihood ratio
test \cite{VanTrees4,Pia2019} whose generic structure is given by
\be
\left[\widehat{\Lambda}_{\hat{i}}(\bZ_L)-c\cdot h(\hat{i})\right] \testest \eta,
\label{eqn:penLRT}
\ee
where
\be
\widehat{i}=\argmax_{i=0,\ldots,M_c}\left\{\widehat{\Lambda}_{i}(\bZ_L)-c\cdot h(i)\right\},
\ee
$\eta$ is the detection threshold to be set according to a desired\footnote{Hereafter, we denote by $\eta$ the generic detection threshold.} 
$P_{fa}$,
\be
\widehat{\Lambda}_{i}(\bZ_L)=
\log\frac{\dmax_{\bR_i}\dmax_{\theta_1,\ldots,\theta_{i}}f_{1,i}(\bZ_L;\widehat{\bM},\bR_i,\bV_i)}{f_0(\bZ_L;\widehat{\bM})}
\label{eqn:GLRT_i}
\ee
with\footnote{Notice that for $i=0$, the maximization at the numerator of (\ref{eqn:GLRT_i}) is with respect to $\sigma^2_{\alpha}$ only. } 
$\widehat{\bM}$ a suitable estimate of $\bM$,
and $c\cdot h(i)$ is a penalty term borrowed from the MOS rules \cite{Stoica1} with $h(i)$ the number of 
unknown real-valued parameters under the $H_{1,i}$ hypothesis and $c$ a suitable scaling
factor. Specifically, $c$ can be set according to the Akaike Information Criterion (AIC), Bayesian Information Criterion (BIC),
and Generalized Information Criterion (GIC), namely
\be
c=
\begin{cases}
1, & \mbox{for AIC-based Detector (AIC-D)},
\\
\log(2LN)/2, & \mbox{for BIC-based Detector (BIC-D)},
\\
(1+\rho)/2, \ \rho>1, & \mbox{for GIC-based Detector (GIC-D)}.
\end{cases}
\ee
Two important remarks are now in order. First, notice that $\widehat{\Lambda}_{i}(\bZ_L)$ 
is the logarithm of a decision statistic obtained by resorting to
a modified Generalized Likelihood Ratio Test (GLRT)-based design procedure \cite{robey1992cfar}, where the ICM is firstly assumed known and then replaced by an estimate.
In the specific case, we use the ML estimate based upon the training samples only 
(namely, the sample covariance matrix that is invertible with probability $1$ since $K\geq N$), 
whose expression is\footnote{This design choice is dictated by the fact that deriving closed-form expression for the plain GLRT is 
not mathematically tractable at least to the best of authors' knowledge.}
\be
\widehat{\bM}=\frac{1}{K}\sum_{k=1}^K\bz_k\bz_k^\dag.
\ee
Second, the correlation of the signals makes the columns of the positive semidefinite matrix $\bR_i$ proportional. Therefore, 
the resulting rank is $1$ and it can be factorized as
\be
\bR_i=\bor_i\bor_i^\dag,
\label{eqn:Rdec}
\ee 
where $\bor_i\in\C^{(i+1)\times 1}$. As a consequence, the number of unknown real-valued parameters under $H_{1,i}$ can be written as
$h(i)=N^2+1+2i$.
With the above remarks in mind, let us proceed by deriving $\widehat{\Lambda}_{0}(\bZ_L)$, namely
\begin{align}
&\widehat{\Lambda}_{0}(\bZ_L)=\log\frac{\dmax_{\sigma_{\alpha}^2}f_{1,0}(\bZ_L;\hbM, \sigma^2_{\alpha},\bv_t)}{f_0(\bZ_L;\hbM)} \nonumber
\\
&=\dmax_{\sigma_{\alpha}^2}\bigg\{\!\!\!-L\log\det(\hbM+\sigma_{\alpha}^2\bv_t\bv_t^\dag) 
-\tr\bigg[\left(\hbM+\sigma_{\alpha}^2\bv_t\bv_t^\dag\right)^{-1} \nonumber
\\
&\times\bZ_L\bZ_L^\dag\bigg]\bigg\} +L\log\det(\hbM)+\tr\left[\hbM^{-1}\bZ_L\bZ_L^\dag\right].
\label{H1 statistic}
\end{align}
Applying the Woodbury identity \cite{MatrixAnalysis}, we come up with the following expression
\be
\widehat{\Lambda}_{0}(\bZ_L)=\dmax_{\sigma_{\alpha}^2}\left[-L\log(1+\sigma_{\alpha}^2\lVert\bv_w\rVert^2)
+\frac{\sigma_{\alpha}^2 \bv_w^\dag\bS_{w}\bv_w }{1+\sigma_{\alpha}^2\lVert\bv_w\rVert^2}\right],
\label{H1 modified}
\ee
where $\bv_w=\hbM^{-1/2}\bv_t$ and $\bS_{w}=\hbM^{-1/2}\bZ_L\bZ_L^\dag\hbM^{-1/2}$.
Let us define the function
\be
g_0(\sigma^2_\alpha)=
-L\log(1+\sigma_{\alpha}^2\lVert\bv_w\rVert^2)
+\frac{\sigma_{\alpha}^2}{1+\sigma_{\alpha}^2\lVert\bv_w\rVert^2}\bv_w^\dag\bS_{w}\bv_w
\ee
and observe that
$\lim_{\sigma^2_\alpha\rightarrow 0} g_0(\sigma^2_\alpha)=0$ and $\lim_{\sigma^2_\alpha\rightarrow +\infty} g_0(\sigma^2_\alpha)=-\infty$.
Thus, in order to maximize $g_0(\sigma^2_\alpha)$, we find the zeros of its first derivative with respect to $\sigma_{\alpha}^2$ to obtain
\be
\widehat{\sigma}_{\alpha}^2=
\begin{cases}
\ds \frac{\bv_w^\dag\bS_{w}\bv_w-L\lVert\bv_w\rVert^2}{L\lVert\bv_w\rVert^4}, \ & \mbox{if} \  \bv_w^\dag\bS_{w}\bv_w>L\lVert\bv_w\rVert^2,
	\\
	0, \ & \mbox{otherwise}.
\end{cases}
\label{eqn:sigma_a_est}
\ee
Moreover, $g_0(\sigma^2_\alpha)$ is monotonic increasing when $\sigma^2_\alpha<({\bv_w^\dag\bS_{w}\bv_w-L\lVert\bv_w\rVert^2})
/({L\lVert\bv_w\rVert^4})$ and decreasing for $\sigma^2_\alpha>({\bv_w^\dag\bS_{w}\bv_w-L\lVert\bv_w\rVert^2})
/({L\lVert\bv_w\rVert^4})$ (provided that $\bv_w^\dag\bS_{w}\bv_w>L\lVert\bv_w\rVert^2$). As a consequence, \eqref{eqn:sigma_a_est} represents
a maximum point of $g_0(\sigma^2_\alpha)$.
It follows that \eqref{H1 modified} can be recast as
\begin{align}\label{eqn:H10finalstatistic}
&\widehat{\Lambda}_{0}(\bZ_L)=\nonumber
\\
&\begin{cases}
	-L\log\!\left(\ds\frac{\bv_w^\dag\bS_{w}\bv_w}{L\lVert\bv_w\rVert^2}\right)
	\!\!+\!\!\ds\frac{\bv_w^\dag\bS_{w}\bv_w}{\lVert\bv_w\rVert^2}\!-\!L,   \!\mbox{if} \  \bv_w^\dag\bS_{w}\bv_w\!>\!L\lVert\bv_w\rVert^2\!\!,
	\\
	0, \  \mbox{otherwise}. 
\end{cases}
\end{align}
The next step towards the computation of \eqref{eqn:penLRT} consists in deriving $\widehat{\Lambda}_{i}(\bZ_L)$ for $i=1,\ldots,M_c$.
To this end, observe that
\begin{align}
&\widehat{\Lambda}_{i}(\bZ_L)=\log\frac{\dmax_{\bR_i}\dmax_{\theta_1,\ldots,\theta_i}
f_{1,i}(\bZ_L;\hbM,\bR_i,\bV_i)}{f_0(\bZ_L;\hbM)} \nonumber
\\
&=\!\!\!\dmax_{{\boldmath R_i}, \atop \theta_1,\ldots,\theta_i} \!\!\!
\bigg\{\!\!\!-L\log\det(\hbM \!+\! \bV_i\bR_i\bV_i^\dag) \!-\! \tr\!\bigg[\!\!\left(\! \hbM \!+\! \bV_i\bR_i\bV_i^\dag\!\right)^{-1}\nonumber
\\
& \times \bZ_L\bZ_L^\dag\bigg]\bigg\} + L\log\det(\hbM)+\tr\left(\hbM^{-1}\bZ_L\bZ_L^\dag\right).
\label{H2 statistic}
\end{align}
Replacing \eqref{eqn:Rdec} into the most right-hand side of the above equation and applying the Woodbury identity leads to
\begin{multline}
\widehat{\Lambda}_{i}(\bZ_L)=\dmax_{\bor_i}\dmax_{\theta_1,\ldots,\theta_i}
\left\{-L\log(1+\bor_i^\dag\bV_{w,i}^\dag\bV_{w,i}\bor_i)\right.\\
\left.+\frac{\bor_i^\dag\bV_{w,i}^\dag\bS_{w}\bV_{w,i}\bor_i}{1+\bor_i^\dag\bV_{w,i}^\dag\bV_{w,i}\bor_i}\right\},
\label{H2 modified}
\end{multline}
where $\bV_{w,i}=\hbM^{-1/2}\bV_i$ and $\bS_w$ has been previously defined. 
Since $\bV_i$ has a Vandermonde structure and $\widehat{\bM}$ is nonsingular, also matrices 
$\bA_i=\bV_{w,i}^\dag\bV_{w,i}$ and $\bB_i=\bV_{w,i}^\dag\bS_{w}\bV_{w,i}$ are nonsingular and \eqref{H2 modified} can be expressed in terms
of these matrices as
\be
\widehat{\Lambda}_{i}(\bZ_L)\!=\!\dmax_{\bor_i}\dmax_{\theta_1,\ldots,\theta_i} 
\left\{\!\!-L\log(1+\bor_i^\dag\bA_i\bor_i)+\frac{\bor_i^\dag\bB_i\bor_i}{1+\bor_i^\dag\bA_i\bor_i}\!\!\right\}.
\label{H2 modified2}
\ee
Exploiting the Cholesky decomposition of $\bA_i$, given by
\be
\bA_i=\bC_i\bC_i^\dag,
\ee
where $\bC_i\in \C^{(i+1)\times (i+1)}$ is a lower triangular matrix, the right-hand side of \eqref{H2 modified2} can be recast as
\begin{align}
& %\widehat{\Lambda}_{i}(\bZ_L)=
\dmax_{\bor_i, \atop \theta_1,\ldots,\theta_i} \!\!\!
\left\{\!\!-L\log(1 \!+\! \bor_i^\dag\bC_i\bC_i^\dag\bor_i)
\!+\! \frac{\bor_i^\dag\bC_i\bC_i^{-1}\bB_i(\bC_i^\dag)^{-1}\bC_i^\dag\bor_i}{1 \!+\! \bor_i^\dag\bC_i\bC_i^\dag\bor_i}\right\} \nonumber
\\
&=\dmax_{\boa_i}\dmax_{\theta_1,\ldots,\theta_i} 
\left\{-L\log(1+\lVert\boa_i\rVert^2)+\frac{\boa_i^\dag\bD_i\boa_i}{1+\lVert\boa_i\rVert^2}\right\},
\label{H2 modified4}
\end{align}
where 
\be
\boa_i=\bC_i^\dag\bor_i, \quad \bD_i=\bC_i^{-1}\bB_i(\bC_i^\dag)^{-1}. 
\label{eqn:matrixDi}
\ee
As further step towards the solution of the above problem,
we decompose $\boa_i$ as the product of a positive scalar $\sqrt{a_i}$, with $a_i>0$, times a unit-norm vector $\bu_{a,i}$. Thus,
problem \eqref{H2 modified4} is tantamount to 
\be
\dmax_{a_i>0}\dmax_{\bu_{a,i} \atop \| u_{a,i}\|=1}\dmax_{\theta_1,\ldots,\theta_i} 
\left\{-L\log(1+a_i) + \frac{a_i {\bu_{a,i}^\dag\bD_i\bu_{a,i}} }{1+a_i}\right\}.
\label{H2 modified5}
\ee
Denoting by 
\be
g(a_i)=-L\log(1+a_i)+ \frac{a_i}{1+a_i}{\bu_{a,i}^\dag\bD_i\bu_{a,i}},
\ee
the maximization with respect to $a_i$ can be performed by investigating the behavior of $g(a_i)$ over the interval $(0,+\infty)$.
Specifically, since the following equalities hold
\be
\lim_{a_i\rightarrow 0} g(a_i)=0, \quad \lim_{a_i\rightarrow +\infty} g(a_i)=-\infty,
\ee
we can search for the stationary points of $g(a_i)$ in the interior of its domain by setting to zero
its first derivative with respect to $a_i$ to obtain
\be
\widehat{a}_i=
\begin{cases}
	\ds\frac{\bu_{a,i}^\dag\bD_i\bu_{a,i}-L}{L},\ & \mbox{if} \ \bu_{a,i}^\dag\bD_i\bu_{a,i}>L,
	\\
	0, \ & \mbox{otherwise}.
\end{cases}
\label{anorm}
\ee
Moreover, it is not difficult to show that $g(a_i)$ is increasing when $a_i<({\bu_{a,i}^\dag\bD_i\bu_{a,i}-L})/{L}$ and decreasing
in the opposite case (provided that $\bu_{a,i}^\dag\bD_i\bu_{a,i}>L$). As a consequence, \eqref{anorm} represents a maximum point.
Now, plugging \eqref{anorm} into \eqref{H2 modified5}, the maximization problem can be recast as 
\eqref{H2 modified6} at the bottom of this page.

The maximization over $\bu_{a,i}$ can be conducted by defining the function 
$f(x)=x-L\log x+L\log L-L$, which is monotonic increasing for $x\geq L$.
In fact, its first derivative is
$\frac{df(x)}{dx}=1-\frac{L}{x}\geq0$, $\forall x\geq L$.
Therefore, the maximum of $f(\bu_{a,i}^\dag\bD_i\bu_{a,i})$ with respect to $\bu_{a,i}$ can be obtained by directly optimizing 
argument, i.e., $\bu_{a,i}^\dag\bD_i\bu_{a,i}$. Exploiting the Rayleigh-Ritz Theorem \cite{MatrixAnalysis}, we obtain that
\be
\dmax_{\bu_{a,i} \atop \| u_{a,i}\|=1} \bu_{a,i}^\dag\bD_i\bu_{a,i}=\lambda_{\max}\{\bD_i\},
\ee
and the maximum is attained when $\bu_{a,i}$ is the normalized eigenvector of $\bD_i$ corresponding to its maximum eigenvalue.
Gathering the above results, we obtain \eqref{eqn:Hifinalstatistic} at the bottom of this page.

\begin{figure}[!tp]
    \centering
    \includegraphics[scale=0.4]{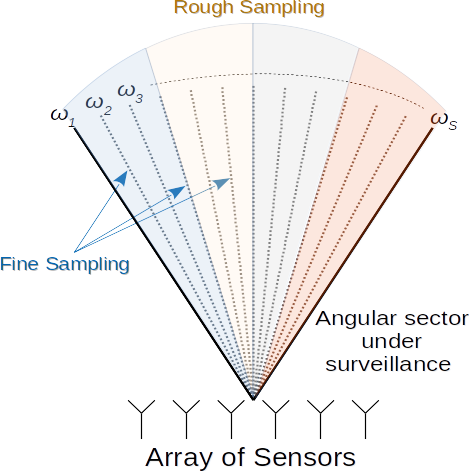}
       \caption{Sampling of the angular sector under surveillance.}
    \label{fig:AngularGrid}
\end{figure}

The last optimization problem to be solved is with respect to the AoAs of the coherent signals. In this respect, since deriving 
closed-form expressions for the ML estimates of $\theta_1,\ldots,\theta_i$ represents a difficult task (at least to
the best of authors' knowledge), we devise grid-search methods. More precisely, the angular sector under surveillance is sampled
to form a discrete set of angular positions which is denoted by $\Omega=\{\omega_1,\ldots,\omega_S\}$ (see Figure \ref{fig:AngularGrid}). 
The cardinality of $\Omega$ can be
chosen according to both the computational power available at the receiver and the system requirements in terms of reactivity 
time intervals. It is clear that
a dense sampling of the angular sector increases the processing time but can provide better estimation results with respect to the case
where the sampling interval is large. 
However, in the latter case, the computational load is lower than in the former case.
Nevertheless, an iterative approach can be pursued for the maximization under the generic $H_{1,i}$ hypothesis and for each 
$i\in\{1,\ldots,M_c\}$. Specifically, given $i\in\{1,\ldots,M_c\}$ and assuming that $H_{1,i}$ holds true, 
we start by roughly sampling the angular sector of interest and use this search grid to come up with
preliminary estimates of the assumed $i$ AoAs. Then, we use these estimates to identify a restricted
angular sector (resulting from the union of partially overlapped narrow sectors) comprising suitable guard 
bands.
\begin{strip}
\begin{equation}
\widehat{\Lambda}_{i}(\bZ_L)=
\begin{cases}
\dmax_{\bu_{a,i} \atop \| u_{a,i}\|=1}\dmax_{\theta_1,\ldots,\theta_i} -L\log(\bu_{a,i}^\dag\bD_i\bu_{a,i})+\bu_{a,i}^\dag\bD_i\bu_{a,i}
+L\log L-L, \ &  \mbox{if} \ \bu_{a,i}^\dag\bD_i\bu_{a,i}>L,
\\
0, \ & \mbox{otherwise}.
\end{cases}
\label{H2 modified6}
\end{equation}
\begin{equation}
\widehat{\Lambda}_{i}(\bZ_L)=
\begin{cases}
	\dmax_{\theta_1,\ldots,\theta_i} -L\log\lambda_{\max}\{\bD_i\}+\lambda_{\max}\{\bD_i\}+L\log L-L, \ & \mbox{if} \  \lambda_{\max}\{\bD_i\}>L,
	\\
	0, \ & \mbox{otherwise}.
\end{cases}
\label{eqn:Hifinalstatistic}
\end{equation}
\end{strip}
The new angular sector of interest is sampled through a shorter sampling interval and the maximization
under $H_{1,i}$ is repeated over the new search grid. 
As a result, the grid points can be very close to the actual angular positions (at least for high
signal power values) improving the estimation quality. Finally, notice that this procedure is applied for each $i=\{1,\ldots,M_c\}$ before 
forming the decision statistic and, in a single application, the number of supposed interferers is maintained constant.
As for the single search procedure under the generic $H_{1,i}$ hypothesis with $i\geq 1$,
observe that in situations where, according to the system geometry and antenna beamwidth, $M_c$ is enough low, 
an exhaustive grid-based search can be conducted over the set
\be
\underbrace{\Omega\times\Omega\times\ldots\times\Omega}_{i \ \mbox{\scriptsize times}}=\Omega^i, \quad \forall i\in\{1,\ldots,M_c\},
\ee
with the constraint that admissible candidates are such that $\omega_1\neq\omega_2\neq\ldots\neq\omega_i$.
On the other hand, exhaustive grid search can be prohibitive for large $M_c$ values. In this case, in order to limit the number of operations,
we design a suboptimum search method relying on the cyclic optimization paradigm \cite{Stoica_alternating}. Specifically, let us
assume that $H_{1,i}$ is true and that $i-1$ estimates, 
$\widehat{\Theta}_{1,i}^{(0)}=\{\hat{\theta}^{(0)}_2,\ldots,\hat{\theta}^{(0)}_i\}\subset\Omega$ say, 
are available, then we select
$\theta_1\in\Omega$ exploiting the following criterion
\begin{multline}
\hat{\theta}_1^{(1)}=\argmax_{\theta_1\in \Omega, \ \theta_1\notin \widehat{\Theta}_{1,i}^{(0)} }\left\{
-L\log\lambda_{\max}\{\bD_i(\theta_1,\widehat{\Theta}_{1,i}^{(0)})\}\right.\\
\left. +\lambda_{\max}\{\bD_i(\theta_1,\widehat{\Theta}_{1,i}^{(0)})\}\right\},
\end{multline}
where $\bD_i(\theta_1,\widehat{\Theta}_1^{(0)})$ is computed as $\bD_i$ in \eqref{eqn:matrixDi} with the difference that
$\bV_i$ is replaced by the following matrix
\be
\bV_i\left(\theta_1,\widehat{\Theta}_{1,i}^{(0)}\right)=\left[
\bv_t,\bv\left(\theta_1\right),\bv(\hat{\theta}_2^{(0)}),\ldots,\bv(\hat{\theta}_i^{(0)})
\right].
\ee
The new estimate $\hat{\theta}_1^{(1)}$ is used to form $\widehat{\Theta}_{2,i}^{(0)}=\{\hat{\theta}^{(1)}_1,
\hat{\theta}^{(0)}_3,\ldots,\hat{\theta}^{(0)}_i\}$ and the update of $\hat{\theta}_2$ is obtained as
\begin{multline}
\hat{\theta}_2^{(1)}=\argmax_{\theta_2\in \Omega, \ \theta_2\notin \widehat{\Theta}_{2,i}^{(0)} }\left\{
-L\log\lambda_{\max}\{\bD_i(\theta_2,\widehat{\Theta}_{2,i}^{(0)})\}\right.\\
\left.+\lambda_{\max}\{\bD_i(\theta_2,\widehat{\Theta}_{2,i}^{(0)})\}\right\},
\end{multline}
where $\bD_i(\theta_2,\widehat{\Theta}_{2,i}^{(0)})$ is built up using
\be
\bV_i\!\!\left(\theta_2,\widehat{\Theta}_{2,i}^{(0)}\right)\!\!=\!\!\left[
\bv_t,\bv(\hat{\theta}_1^{(1)}),\bv(\theta_2),\bv(\hat{\theta}_3^{(0)}),\ldots,\bv(\hat{\theta}_i^{(0)})
\right].
\ee
The above steps continue until the $i$th update occurs to obtain 
$\widehat{\Theta}_{1,i}^{(1)}=\{\hat{\theta}^{(1)}_2,\ldots,\hat{\theta}^{(1)}_i\}$, which can be used to repeat the entire 
refinement procedure. 

Summarizing, the update of the $k$th AoA within the $n$th ($n\geq 1$) procedure cycle has the following expression
\begin{multline}
\hat{\theta}_k^{(n)}=\argmax_{\theta_k\in \Omega, \ \theta_k\notin \widehat{\Theta}_{k,i}^{(n-1)} }\left\{
-L\log\lambda_{\max}\{\bD_i(\theta_k,\widehat{\Theta}_{k,i}^{(n-1)})\}\right.\\
\left.+\lambda_{\max}\{\bD_i(\theta_k,\widehat{\Theta}_{k,i}^{(n-1)})\}\right\},
\end{multline}
$1\leq k\leq i$, where 
\be
\widehat{\Theta}_{k,i}^{(n-1)}=\left\{
\hat{\theta}_1^{(n)},\ldots\hat{\theta}_{k-1}^{(n)},\hat{\theta}_{k+1}^{(n-1)},\ldots,\hat{\theta}_i^{(n-1)}
\right\}
\ee
and $\bD_i(\theta_k,\widehat{\Theta}_{k,i}^{(n-1)})$ is defined as in \eqref{eqn:matrixDi} with
$
\bV_i\left(\theta_k,\widehat{\Theta}_{k,i}^{(n-1)}\right)=[
\bv_t,\bv(\hat{\theta}_1^{(n)}),\ldots,\bv(\hat{\theta}_{k-1}^{(n)}),
\bv(\theta_{k}),
\bv(\hat{\theta}_{k+1}^{(n-1)}),\ldots,\bv(\hat{\theta}_i^{(n-1)})].
$
It is important to observe that the above procedure leads to a nondecreasing sequence
of log-likelihood function values, namely
\begin{align}
&\cL_i(\hat{\theta}_1^{(0)},\widehat{\Theta}_{1,i}^{(0)})\leq \cL_i(\hat{\theta}_1^{(1)},\widehat{\Theta}_{1,i}^{(0)})
=\cL_i(\hat{\theta}_2^{(0)},\widehat{\Theta}_{2,i}^{(0)})
\leq \nonumber
\\
&\cL_i(\hat{\theta}_2^{(1)},\widehat{\Theta}_{2,i}^{(0)})
\leq\ldots\leq
\cL_i(\hat{\theta}_k^{(n)},\widehat{\Theta}_{k,i}^{(n-1)})\leq\ldots
\end{align}
where 
\begin{multline}
\cL_i(\hat{\theta}^{(n)}_k,\widehat{\Theta}^{(m)}_{k,i})=-L\log\lambda_{\max}\{\bD_i(\hat{\theta}^{(n)}_k,\widehat{\Theta}_{k,i}^{(m)})\}
\\+\lambda_{\max}\{\bD_i(\hat{\theta}^{(n)}_k,\widehat{\Theta}_{k,i}^{(m)})\}+L\log L-L.
\end{multline}
The entire procedure may terminate when
\be
\Delta\cL_i(n)=\frac{\left|\cL_i(\hat{\theta}_1^{(n+1)},\widehat{\Theta}_{1,i}^{(n+1)}) 
-\cL_i(\hat{\theta}_1^{(n)},\widehat{\Theta}_{1,i}^{(n)})\right|}
{\left|\cL_i(\hat{\theta}_1^{(n+1)},\widehat{\Theta}_{1,i}^{(n+1)})\right|}<\epsilon,
\ee
where $\epsilon>0$, or $n\geq N_{\max}$ where $N_{\max}$ is the maximum allowable number of iterations.

Finally, as for the initialization of the procedure, given the received vectors, we proceed by first computing
\be
r_1=\left|\frac{1}{L}\sum_{l=1}^L \bz_l^\dag\bv(\omega_1)\right|^2, \ldots, r_S=\left|\frac{1}{L}\sum_{l=1}^L \bz_l^\dag\bv(\omega_S)\right|^2,
\ee
then, sort the above values in decreasing order, namely
\be
r_{k_1} \geq r_{k_2} \geq \ldots \geq r_{k_S},
\ee
and select $\omega_{k_1},\ldots,\omega_{k_i}$.
The behavior of the proposed architectures coupled with the above estimation procedure is assessed in 
the next section by means of numerical examples.

\section{Illustrative Examples and Disussion}
\label{Sec:Performance}
In this section, the performance of the proposed detection architectures is investigated
drawing upon synthetic data and considering three operating scenarios. In the first case, only the target of interest is present,
whereas in the second scenario, an additional coherent signal is considered. Finally, the third scenario contains the signal of interest
along with two additional coherent signals. The performance metrics are
\begin{itemize}
\item the Probability of Detection under $H_{1,i}$, $i\geq 0$, ($P_{d,i}$) defined as the probability of rejecting $H_0$ 
when the latter is false and given a preassigned $P_{fa}=P(\mbox{reject } H_0|H_0 \mbox{ is true})$;
\item the Probability of Correct Classification ($P_{cc}$), namely the probability of deciding\footnote{The detection threshold is 
set according to the preassigned $P_{fa}$ also in this case.} 
for $H_{1,i}$ under $H_{1,i}$;
\item the Root Mean Square Error (RMSE) in angle whose expression is
\be
\sqrt{\E\left[\frac{1}{M}\sum_{m=1}^M \dmin_{k=1,\ldots,\hat{i}}\left\{ (\theta_m - \hat{\theta}_k)^2 \right\}\right] },
\label{eqn:RMSE_def}
\ee
\end{itemize}
where $\hat{\theta}_k$, $k=1,\ldots,\hat{i}$, are the AoA estimates of the coherent signals when $H_{1,\hat{i}}$ is declared.
Since deriving closed-form expressions for the above quantities represents a mathematically intractable task (at least to the best
of authors' knowledge), we estimate them resorting to the Monte Carlo counting techniques (also replacing the 
statistical expectation of the RMSE with the sample mean over the Monte Carlo trials). More precisely, we exploit
$10^3$ independent trials to estimate the considered performance metrics, whereas the  detection thresholds are set 
over $100/P_{fa}$, with $P_{fa}=10^{-3}$.

As stated before, the analysis starts from the conventional case where only the target echoes impinge on the radar 
and proceeds with two more difficult cases
that assume the presence of two and three coherent signals (including that of interest), respectively. 
As for the operating setup, the coherent signals share the same power as the target 
signal, whereas the angular sector under surveillance ranges from $-25$ to $25$ degrees and is sampled at $1$ degree. 
The Signal-to-Interference-plus-Noise Ratio (SINR) is defined as
$\mbox{SINR}=\sigma_\alpha^2\bv_t^{\dagger}\bM^{-1}\bv_t$,
where $\bM=\sigma_n^2\bI+\mbox{CNR} \, \bM_c$, $\mbox{CNR}=20$ dB is the Clutter-to-Noise Ratio,
$\sigma^2_n=1$ is the noise power, and $\bM_c$ is the clutter covariance matrix whose $(i,j)$th entry is defined as 
$\bM_c(i,j)=\rho_c^{|i-j|}$ with $\rho_c=0.9$ the one-lag correlation coefficient. 
Moreover, all the numerical examples assume $N=16$, $L=32$, $\theta_t=0^{\circ}$, and $M_c=4$.

As for the angular positions of the coherent signals, we begin the analysis assuming that
they belong to the search grid. In this case, the obtained curves represent an upper bound on the performance that
can be attained exploiting fine and fine search grids (possibly considering the previously described iterative maximization procedure).
Then, we investigate the behavior of the proposed architectures when the actual positions of the coherent signals
are in between the points of the search grid. Specifically, we generate them as uniformly distributed in intervals of
different sizes and centered on the nominal search grid points (a point better explained below).

Finally, we compare the proposed architectures with well-known decision schemes, namely
the Generalized Adaptive Matched Filter (GAMF) and the Generalized Adaptive Subspace Detector (GASD) given by \cite{GLRT-based}
\begin{align}
&\sum_{l=1}^L\frac{|\bv_t^\dag\widehat{\bM}^{-1}\bz_l|^2}{\bv_t^\dag\widehat{\bM}^{-1}\bv_t}\testconv \eta   \quad \mbox{and}
\\ 
&\sum_{l=1}^L\frac{|\bv_t^\dag\widehat{\bM}^{-1}\bz_l|^2}{\bv_t^\dag\widehat{\bM}^{-1}\bv_t \sum_{m=1}^L \bz_m^\dag\widehat{\bM}^{-1}\bz_m}
\testconv \eta,
\end{align}
respectively.

\begin{figure}[htbp]
	\centering
	\includegraphics[width=0.4\textwidth] {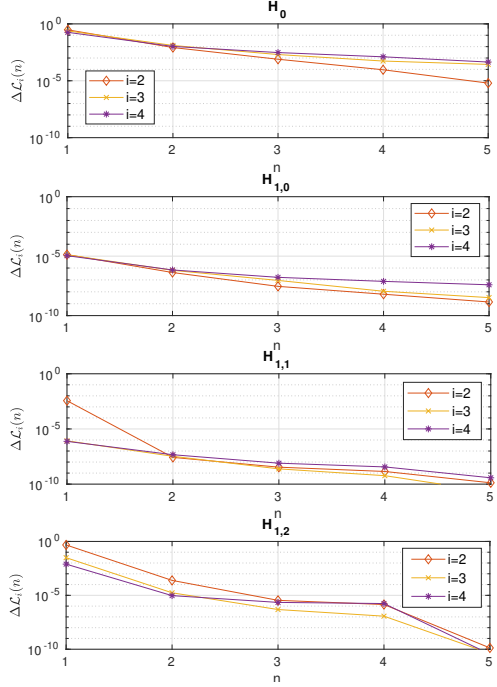}
	\caption{Convergence curves under $H_0$, $H_{1,0}$, $H_{1,1}$, and $H_{1,2}$ assuming $N=16$, $L=32$, $K=32$, 
	$\theta_1=10^\circ$, and $\theta_2=18^\circ$.}
	\label{fig:convergenceAnalysis_powerRanking}
\end{figure}

\begin{figure}[htbp]
	\centering
	\includegraphics[width=0.38\textwidth] {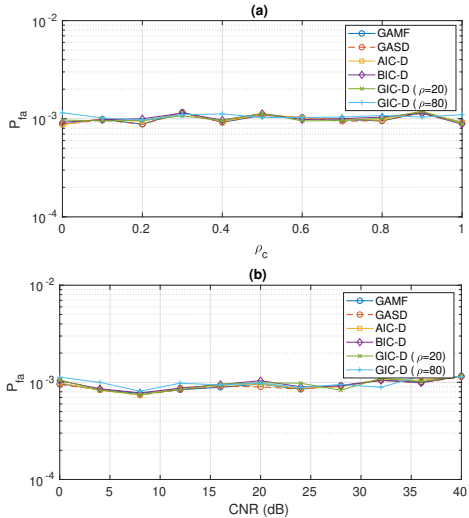}
	\caption{$P_{fa}$ versus $\rho_c$ (subplot (a)) and CNR (subplot (b)) for the GAMF, GASD, AIC-D, BIC-D, and GIC-D 
	with different values of $\rho$ assuming $N=16$, $L=32$, and $K=32$.}
	\label{fig:pfa_CNR_rho}
\end{figure}

Before presenting the detection and classification results for each of the aforementioned cases, 
it is important to assess the convergence rate of the estimation procedure under each hypothesis 
as well as the sensitivity of the $P_{fa}$ with respect to $\rho_c$ and CNR. To this end, in 
Figure \ref{fig:convergenceAnalysis_powerRanking}, we plot the root mean square values for $\Delta\cL_i(n)$, $i=2,3,4$, versus $n$ 
under $H_0$, $H_{1,0}$, $H_{1,1}$, and $H_{1,2}$; as for $H_{1,1}$ and $H_{1,2}$, we set $\theta_1=10^\circ$ and $\theta_2=18^\circ$. 
Inspection of the figure highlights that a number of iterations $n=5$ is enough to guarantee $\epsilon\leq 10^{-3}$ under each hypothesis.
As for the $P_{fa}$ behavior, in Figure \ref{fig:pfa_CNR_rho}, we estimate it for different values of $\rho_c$ given 
CNR (subplot (a)) and 
for different values of CNR given $\rho_c$ (subplot (b)) when the thresholds are computed assuming 
the nominal values for these parameters (namely, $\rho_c=0.9$ and $\mbox{CNR}=20$ dB) and $P_{fa}=10^{-3}$.
It turns out that all the considered architectures can guarantee $P_{fa}$ values contained within the interval $[0.0007, \ 0.0011]$ 
providing a rather robust behavior to the considered parameter variations.


\begin{figure}[tbp]
	\centering
	\includegraphics[width=0.4\textwidth] {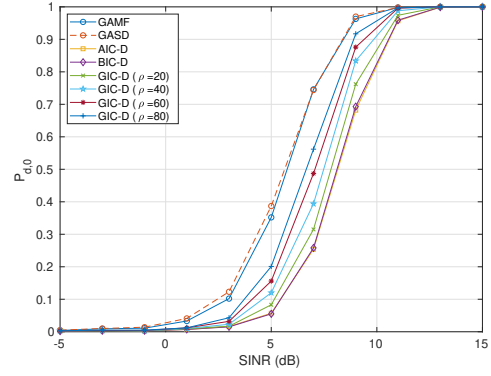}
	\caption{$P_{d,0}$ versus SINR for the GAMF, GASD, AIC-D, BIC-D, and GIC-D with different values of $\rho$ assuming $N=16$, $L=32$, and $K=32$.}
	\label{fig:Pd_0vsSINR_K32}
\end{figure}

\begin{figure}[tbp]
	\centering
	\includegraphics[width=0.42\textwidth] {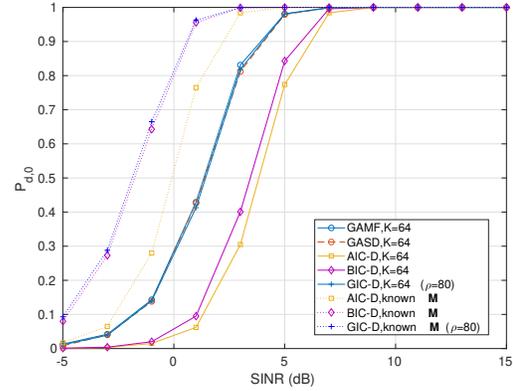}
	\caption{$P_{d,0}$ versus SINR for the GAMF, GASD, AIC-D, BIC-D, GIC-D with $\rho=80$, AIC-D with known $\bM$, BIC-D with known $\bM$, and 
	GIC-D with known $\bM$ and $\rho=80$ assuming $N=16$, $L=32$, and $K=64$.}
	\label{fig:Pd_0vsSINR_K64}
\end{figure}

\begin{figure}[tbp]
	\centering
	\includegraphics[width=0.5\textwidth] {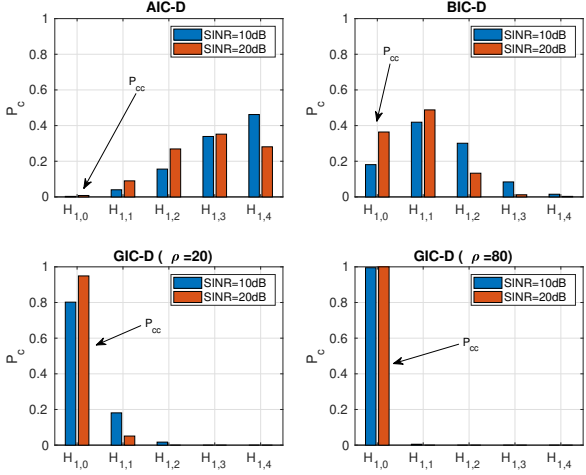}
	\caption{Classification probabilities for the AIC-D, BIC-D, GIC-D with $\rho=20$, and GIC-D with $\rho=80$ under $H_{1,0}$ assuming $N=16$, $L=32$, and $K=32$.}
	\label{fig:class_H_10}
\end{figure}

In the next three figures, we investigate the detection and classification performance when the conventional alternative hypothesis
$H_{1,0}$, which contemplates the presence of the signal of interest only (first scenario), is in force. 
Specifically, in Figure \ref{fig:Pd_0vsSINR_K32}, we show the $P_{d,0}$ curves as functions of 
the SINR assuming $N=16$, $L=32$, $K=32$, and different values for $\rho$. This preliminary analysis allows us to quantify the 
sensitivity of GIC-D with respect to its tuning parameter in comparison with the other
considered architectures. It turns out that increasing $\rho$ leads to improved detection performances for GIC-D, whose
loss with respect to both the GAMF and GASD, which overcome the other architectures,
ranges from about $1.9$ dB for $\rho=20$ to about $0.5$ dB for $\rho=80$ at $P_{d,0}=0.9$. On the other hand, AIC-D and BIC-D experience a 
loss of about $2.2$ dB with respect to GAMF and GASD at $P_{d,0}=0.9$. In Figure \ref{fig:Pd_0vsSINR_K64}, we analyze the 
effects of $K$ on the performance by doubling the value used in Figure \ref{fig:Pd_0vsSINR_K32}. 
Moreover, in order to quantify the loss due to the estimation of $\bM$
we also report the curves of the clairvoyant detectors which assume that $\bM$ is known. The most evident change with respect to the previous
figure is that all the detection curves clearly move towards the left part of the plot, namely a significant improvement in performance
occurs. In addition, GIC-D with $\rho=80$, GAMF, and GASD almost share the same performance while AIC-D and BIC-D continue to exhibit a loss
of about $2$ dB at $P_{d,0}=0.9$ with respect to the former. The figure also highlights that the loss associated 
to the estimation of $\bM$ is of about $3.5$ dB (at $P_{d,0}=0.9$). In the last figure of this first study case (Figure \ref{fig:class_H_10}), 
we analyze the classification capabilities of the proposed architectures through the probabilities of 
classification ($P_c$) and the $P_{cc}$ 
estimated at two different SINR values.
The figure clearly highlights the inclination of AIC-D to overestimate the hypothesis model, whereas the penalty term of BIC-D allows
to mitigate this effect even though the $P_{cc}$ is less than $0.4$. On the other hand, GIC-D can be suitably tuned in order to
attain satisfactory classification performance. Specifically, when $\rho=80$, GIC-D correctly classifies the environment with a probability
very close to $1$.

\begin{figure}[htbp]
	\centering
	\includegraphics[width=0.41\textwidth] {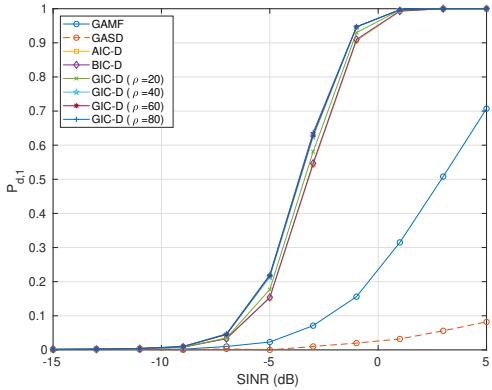}
	\caption{$P_{d,1}$ versus SINR for the GAMF, GASD, AIC-D, BIC-D, and GIC-D with different values of $\rho$ assuming 
	$N=16$, $L=32$, $K=32$, and $\theta_1=10^\circ$.}
	\label{fig:Pd_1vsSINR_K32}
\end{figure}
\begin{figure}[htbp]
	\centering
	\includegraphics[width=0.42\textwidth] {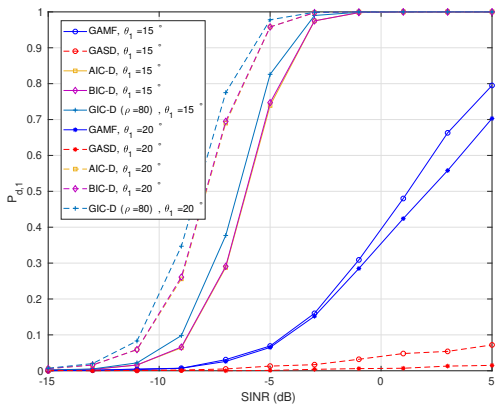}
	\caption{$P_{d,1}$ versus SINR for the GAMF, GASD, AIC-D, BIC-D, and GIC-D assuming different values of $\theta_1$, 
	$N=16$, $L=32$, and $K=32$.}
	\label{fig:Pd_1vsSINR_K32_theta1}
\end{figure}
\begin{figure}[htbp]
	\centering
	\includegraphics[width=0.42\textwidth] {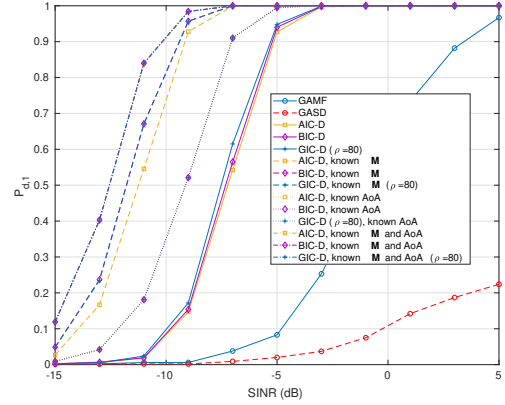}
	\caption{$P_{d,1}$ versus SINR for the GAMF, GASD, AIC-D, AIC-D with known parameters, AIC-D with known $\bM$ only, AIC-D with known AoA only,
	BIC-D, BIC-D with known parameters, BIC-D with known $\bM$ only, BIC-D with known AoA only, GIC-D with $\rho=80$, GIC-D with $\rho=80$ and
	known parameters, GIC-D with $\rho=80$ and known $\bM$ only, and GIC-D with $\rho=80$ and known AoA only, assuming $N=16$, $L=32$, $K=64$,
	and $\theta_1=10^\circ$.}
	\label{fig:Pd_1vsSINR_K64}
\end{figure}
\begin{figure}[htbp]
	\centering
	\includegraphics[width=0.5\textwidth] {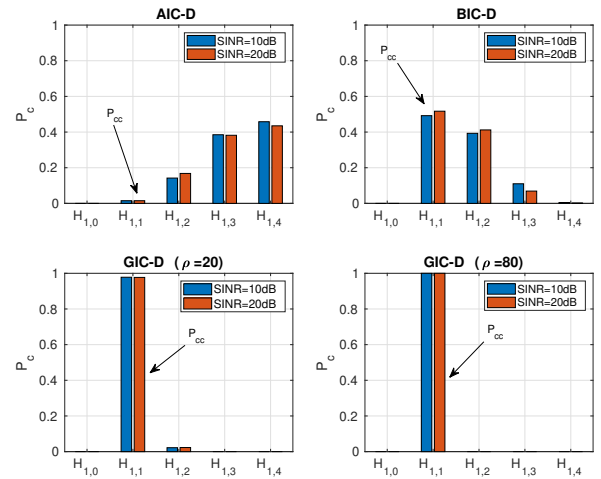}
	\caption{Classification probabilities for the AIC-D, BIC-D, and GIC-D with $\rho=20,80$ under $H_{1,1}$ assuming $N=16$, $L=32$, $K=32$,
	and $\theta_1=10^\circ$.}
	\label{fig:class_H_11}
\end{figure}
\begin{figure}[htbp]
	\centering
	\includegraphics[width=0.43\textwidth] {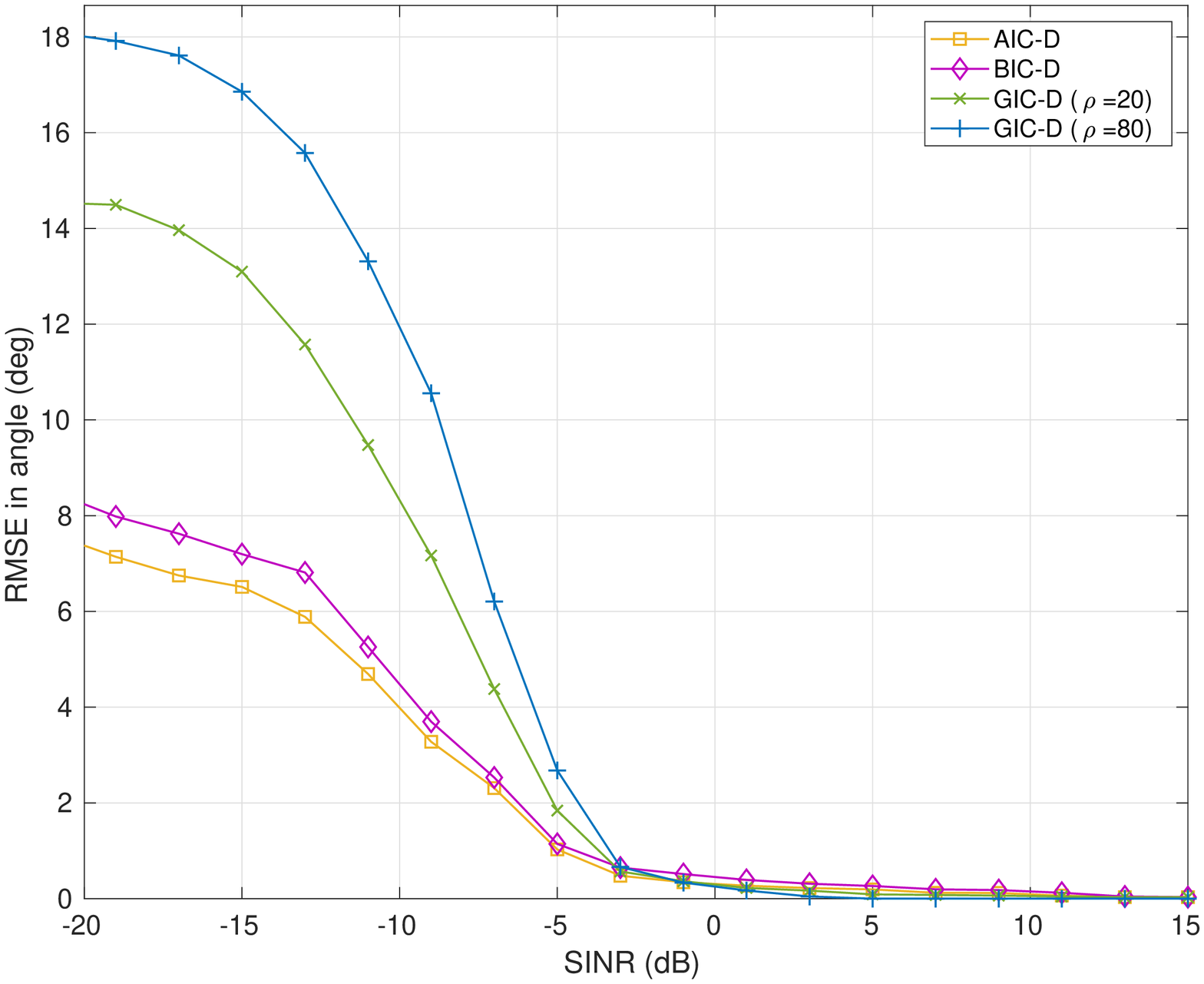}
	\caption{RMSE in angle versus SINR for the AIC-D, BIC-D, and GIC-D with $\rho=20,80$ under $H_{1,1}$ 
	assuming $N=16$, $L=32$, $K=32$, and $\theta_1=10^\circ$.}
	\label{fig:RMSE_H_11}
\end{figure}
Now, we address the intermediate scenario, where besides the signal of interest, a coherent signal enters the antenna from an angular
direction $\theta_1$. The hypothesis corresponding to this scenario is $H_{1,1}$. In Figure \ref{fig:Pd_1vsSINR_K32}, 
we plot $P_{d,1}$ versus SINR for $N=16$, $L=32$, $K=32$, and $\theta_1=10^\circ$; moreover, several values of $\rho$ are considered. 
Notice that in the presence of the additional coherent signal, GAMF and GASD are no longer at the top of the 
performance ranking with the GASD not capable of providing a $P_{d,1}$ above $0.1$ due to its selective behavior \cite{BOR-Morgan}, 
at least for the considered parameter values. 
On the other hand, the remaining detectors share about the same performance with a gain of about $10$ dB with respect to the GAMF; 
observe also that GIC-D slightly improves its performance as $\rho$ grows. In the next figure, we assess the behavior of the considered
decision schemes when $\theta_1$ takes on different values. Specifically, we set $\theta_1=15^\circ, 20^\circ$, whereas the remaining
parameters are the same as in the previous figure except for $\rho=80$. The figure shows that the performance of the proposed 
architectures exhibits a gain of about $2$ dB (at $P_{d,1}=0.9$) when the angular separation between the coherent signals increases with the
GIC-D slightly outperforming the other schemes, whereas the performance of the GAMF and GASD degrades due to the fact that
the composition of the coherent signals results in a direction that moves away from the nominal steering angle. In Figure \ref{fig:Pd_1vsSINR_K64}, 
we account for a different value of $K$ leaving unaltered the parameters $N,L$ and also quantify the loss due to the estimation of 
the AoA and/or $\bM$. As expected, increasing $K$ improves the performance of all detectors. Moreover,
it turns out that estimating the AoA is not as crucial as the estimation of the covariance matrix since the loss associated
with the estimation of AoA is of about $1$ dB at $P_{d,1}=0.9$ and with respect to the architectures where all the parameters
are known, by contrast that associated with the estimation of $\bM$ is approximately $3$ dB. Again, GIC-D with $\rho=80$ 
slightly overcomes the other proposed schemes. In the last two figures related to this intermediate scenario, we show the classification
and estimation performance. Specifically, Figure \ref{fig:class_H_11} contains the classification histograms, whereas 
the curves of RMSE in angle versus the SINR are presented in Figure \ref{fig:RMSE_H_11}. The former figure confirms the behavior observed in
Figure \ref{fig:class_H_10} and, hence, the superior performance of GIC-D with respect to AIC-D and BIC-D that are inclined
to overestimate the number of coherent signals. 
As for the RMSE, all the proposed architectures share almost the same performance
for SINR$\geq -5$ dB. 
The main differences occur for low SINR values where, based upon \eqref{eqn:RMSE_def}, the overestimation
of the number of coherent signals for the AIC-D and BIC-D leads to lower RMSE values than GIC-D.
As a matter of fact, given \eqref{eqn:RMSE_def} and for low SINR values, it is more likely to draw an estimate
from the set provided by AIC-D/BIC-D that is in a narrower neighborhood of a true angular position with respect 
to all the estimates provided by GIC-D. As a consequence, the resulting
error for AIC-D/BIC-D takes on lower values than the error for GIC-D. 

\begin{figure}[t!]
	\centering
	\includegraphics[width=9cm, height=7.2cm] {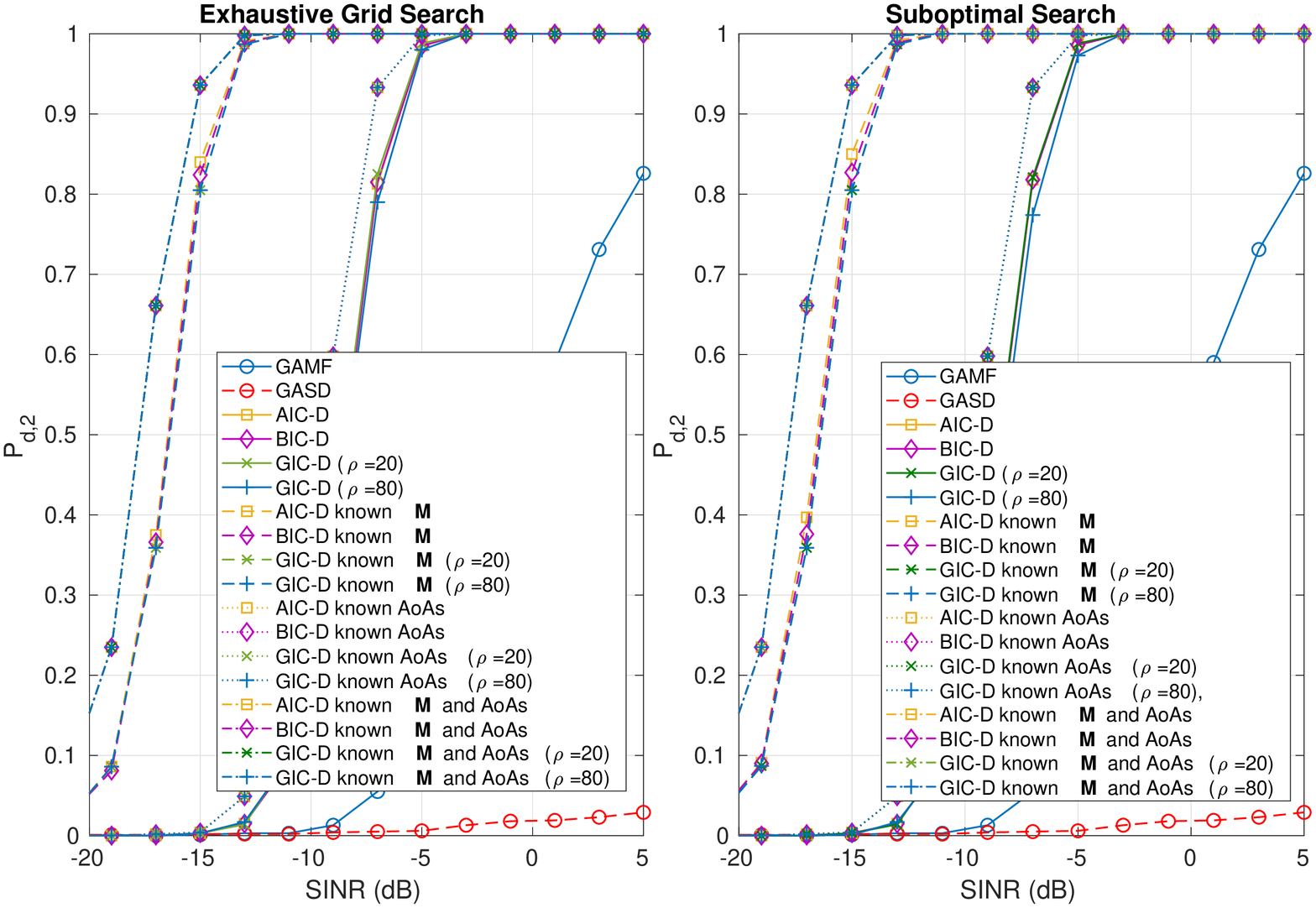}
	\caption{$P_{d,2}$ versus SINR for the GAMF, GASD, AIC-D, AIC-D with known parameters, AIC-D with known $\bM$ only, AIC-D with known AoA only,
	BIC-D, BIC-D with known parameters, BIC-D with known $\bM$ only, BIC-D with known AoA only, GIC-D with $\rho=20,80$, GIC-D with $\rho=20,80$ 
	and	known parameters, GIC-D with $\rho=20,80$ and known $\bM$ only, and GIC-D with $\rho=20,80$ and known AoA only, 
	exhaustive grid search (left subplot) and suboptimum search (right subplot) assuming $N=16$, $L=32$, $K=32$, $\theta_1=10^\circ$, 
	and $\theta_2=18^\circ$.}
	\label{fig:Pd_2vsSINR_K32}
\end{figure}

\begin{figure}[tbp]
	\centering
	\includegraphics[width=0.5\textwidth] {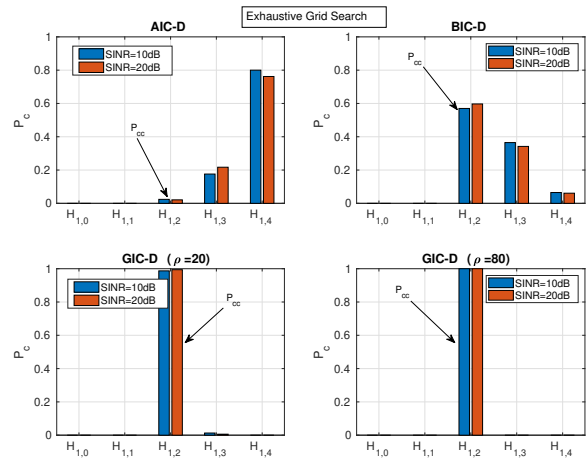}
	\caption{Classification probabilities for the AIC-D, BIC-D, and GIC-D with $\rho=20,80$ under $H_{1,2}$ assuming $N=16$, $L=32$, $K=32$,
	$\theta_1=10^\circ$, $\theta_2=18^\circ$, and the exhaustive grid search.}
	\label{fig:class_H_12_exhaustive}
\end{figure}

\begin{figure}[tbp]
	\centering
	\includegraphics[width=0.5\textwidth] {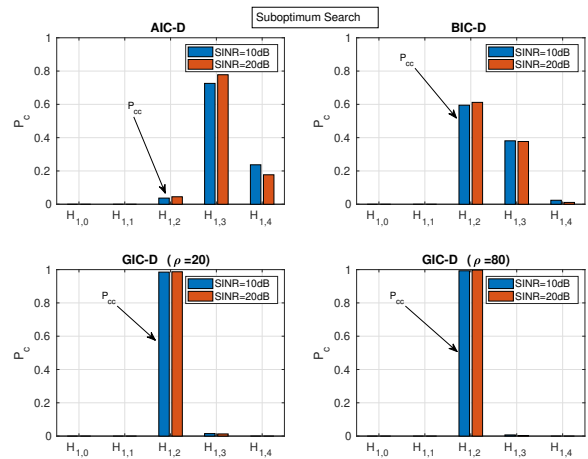}
	\caption{Classification probabilities for the AIC-D, BIC-D, and GIC-D with $\rho=20,80$ under $H_{1,2}$ assuming $N=16$, $L=32$, $K=32$,
	$\theta_1=10^\circ$, $\theta_2=18^\circ$, and the suboptimum search.}
	\label{fig:class_H_12_suboptimum}
\end{figure}

\begin{figure}[tbp]
	\centering
	\includegraphics[width=0.5\textwidth] {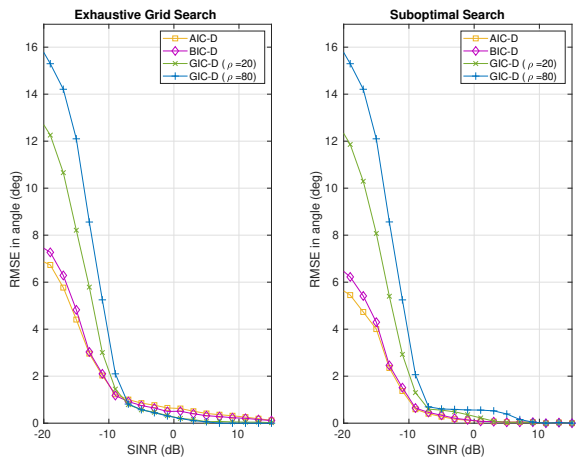}
	\caption{RMSE in angle versus SINR for the AIC-D, BIC-D, and GIC-D with $\rho=20,80$ under $H_{1,2}$ 
	assuming $N=16$, $L=32$, $K=32$, $\theta_1=10^\circ$, $\theta_2=18^\circ$, exhaustive search grid (left subplot), and
	suboptimum search (right subplot).}
	\label{fig:RMSE_H_12}
\end{figure}

\begin{figure}[tbp]
 	\centering
 	\includegraphics[width=0.45\textwidth] {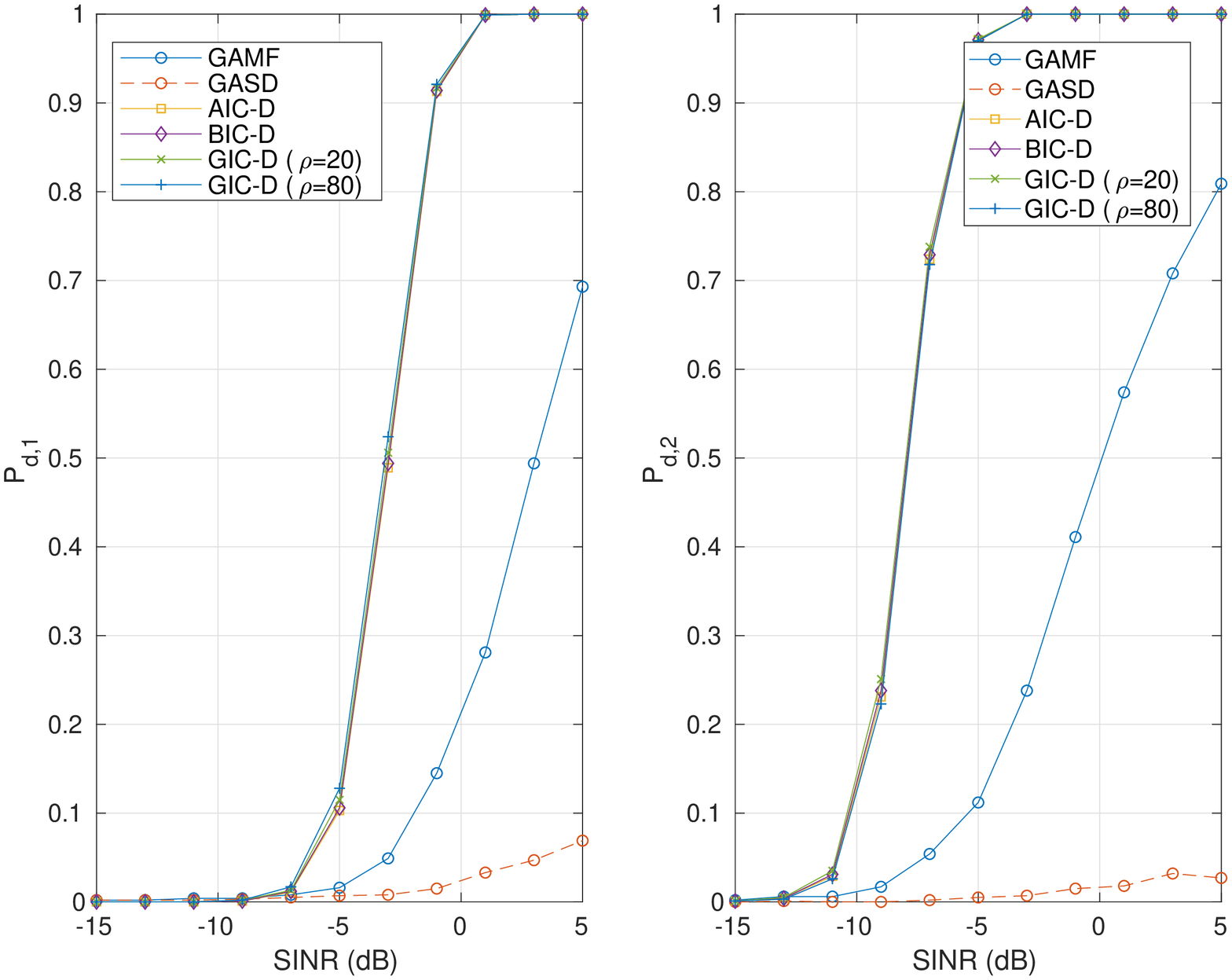}
 	\caption{$P_{d,1}$ versus SINR (left subplot) and $P_{d,2}$ versus SINR (right subplot) for the GAMF, GASD, AIC-D, BIC-D, and GIC-D
 	with $\rho=20,80$ when the actual AoAs of the coherent signals are uniformly generated in between $\theta_i-0.3^\circ$ and
 	$\theta_i+0.3^\circ$, $i=1,2$, assuming $N = 16$, $L = 32$, $K = 32$, $\theta_1 = 10^{\circ}$, $\theta_2 = 18^{\circ}$,
 	and the suboptimum search.}
	\label{fig:pdMis03}
\end{figure}

\begin{figure}[tbp]
 	\centering
 	\includegraphics[width=0.45\textwidth] {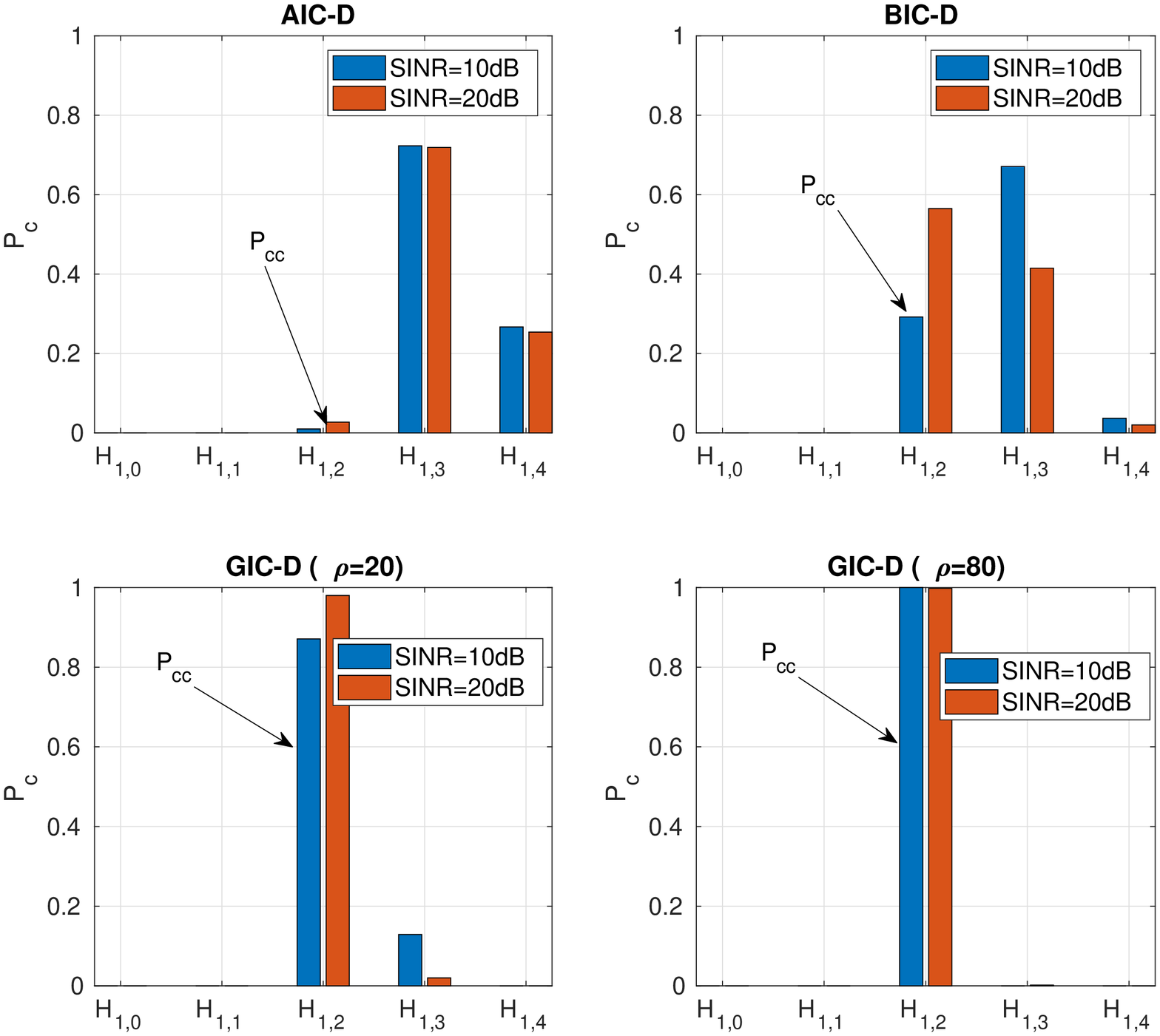}
 	\caption{Classification probabilities for AIC-D, BIC-D, GIC-D with $\rho= 20,80$ under $H_{1,2}$ when the actual AoAs 
 	of the coherent signals are uniformly generated in between $\theta_i-0.3^\circ$ and
 	$\theta_i+0.3^\circ$, $i=1,2$, assuming $N = 16$, $L = 32$, $K = 32$, $\theta_1 = 10^{\circ}$, $\theta_2 = 18^{\circ}$,
 	and the suboptimum search.}
 	\label{fig:classMis_03}
\end{figure}

\begin{figure}[tbp]
 	\centering
 	\includegraphics[width=0.45\textwidth] {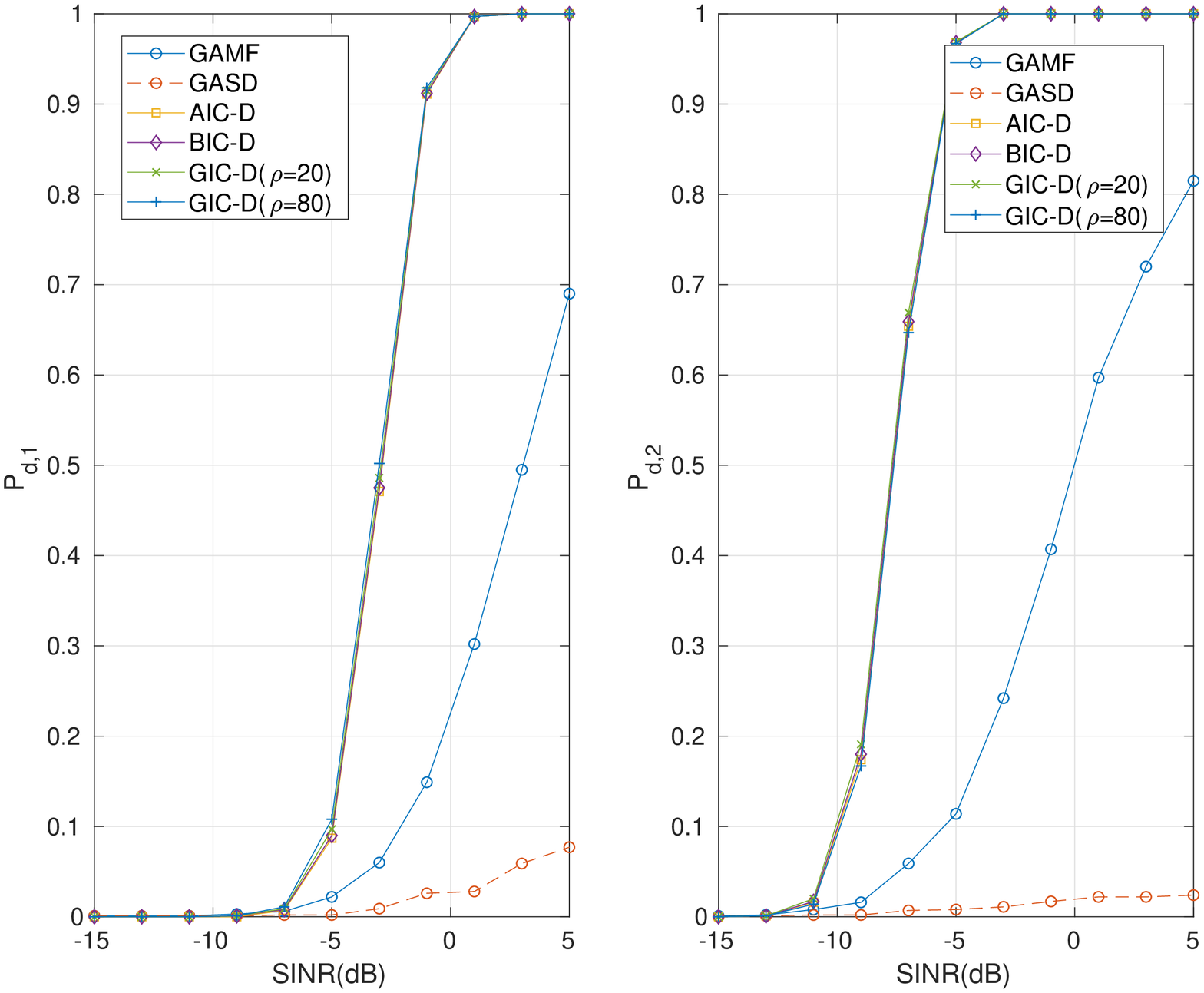}
 	\caption{$P_{d,1}$ versus SINR (left subplot) and $P_{d,2}$ versus SINR (right subplot) for the GAMF, GASD, AIC-D, BIC-D, and GIC-D
 	with $\rho=20,80$ when the actual AoAs of the coherent signals are uniformly generated in between $\theta_i-0.5^\circ$ and
 	$\theta_i+0.5^\circ$, $i=1,2$, assuming $N = 16$, $L = 32$, $K = 32$, $\theta_1 = 10^{\circ}$, $\theta_2 = 18^{\circ}$,
 	and the suboptimum search.}
 	\label{fig:pdMis05}
\end{figure}

\begin{figure}[tbp]
 	\centering
 	\includegraphics[width=0.45\textwidth] {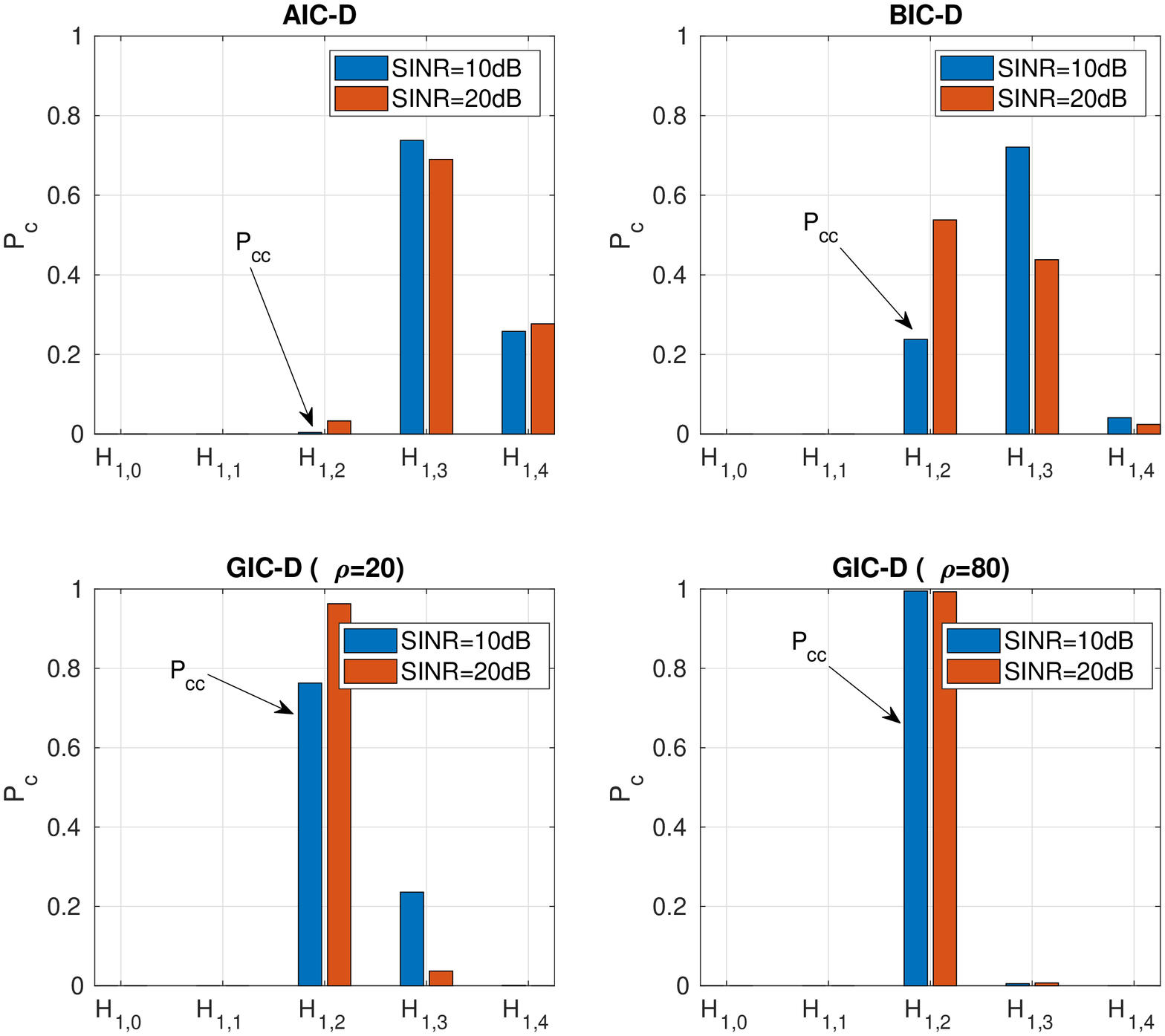}
 	\caption{Classification probabilities for AIC-D, BIC-D, GIC-D with $\rho= 20,80$ under $H_{1,2}$ when the actual AoAs 
 	of the coherent signals are uniformly generated in between $\theta_i-0.5^\circ$ and
 	$\theta_i+0.5^\circ$, $i=1,2$, assuming $N = 16$, $L = 32$, $K = 32$, $\theta_1 = 10^{\circ}$, $\theta_2 = 18^{\circ}$,
 	and the suboptimum search.}
 	\label{fig:classMis_05}
\end{figure}

Finally, the last scenario is the most challenging since it encompasses the presence of two 
coherent signals with nominal positions $\theta_1=10^\circ$ 
and $\theta_2=18^\circ$ in addition to the signal 
of interest. Besides, we consider two situations that differ in the actual positions of the coherent signals.
In the first situation, they are exactly located at $\theta_1$ and $\theta_2$, whereas in the second case,
the positions of the two coherent signals are uniformly generated in the intervals $[\theta_1-\Delta\theta, \ \theta_1+\Delta\theta]$
and $[\theta_2-\Delta\theta, \ \theta_2+\Delta\theta]$, where $\Delta\theta\in\{0.3^\circ,0.5^\circ\}$.
The detection performance for matched signals is shown in Figure \ref{fig:Pd_2vsSINR_K32}, where we also plot the architectures that 
assume partial/full knowledge of the parameter values. Moreover, we compare the results obtained through the exhaustive search 
for the AoA estimation (left subplot) with those provided by the proposed suboptimum procedure (right subplot). 
The figure highlights that the search procedures return detection curves
that are very close to each other. In addition, the same remarks for Figure \ref{fig:Pd_1vsSINR_K64} also hold in this case.
The classification performance and the RMSE curves for both the exhaustive and suboptimum search procedures are shown 
in Figure \ref{fig:class_H_12_exhaustive}-\ref{fig:RMSE_H_12}. Inspection of these figures confirms the excellent 
classification capabilities of GIC-D, whereas AIC-D and BIC-D can provide better AoA estimates than GIC-D for low SINR values ($\leq -10$ dB).
In addition, it is important to underline that there does not exist a valuable difference in performance between the
exhaustive search and the suboptimum search algorithm.
Finally, the last four figures assess the behavior of the considered architectures 
(coupled with the suboptimum search procedure)
when the AoAs of the coherent signals
are uniformly generated at each Monte Carlo trial in an interval of length $2\Delta\theta$ and centered around the nominal positions.
Figures \ref{fig:pdMis03} and \ref{fig:classMis_03} assume $\Delta\theta=0.3$, whereas 
in Figures \ref{fig:pdMis05} and \ref{fig:classMis_05}, we set $\Delta\theta=0.5$. 
These numerical examples highlight
that, from the detection point of view, the mismatch between the search grid points and the actual positions of the coherent signals
leads to a slight performance deterioration, that is more noticeable for $\Delta\theta=0.5$, while leaving the previously
observed hierarchy unaltered. Note that we do not assume any mismatch related to the signal of interest since it requires a different 
performance analysis that accounts for the mismatch degree and is out of the scope of the present work.
As for the classification performance, AIC-D and BIC-D continue to exhibit a clear inclination to overestimate
the number of signals, while GIC-D with $\rho=20$ returns a probability of correct classification lower than that for matched signals.
Finally, the classification performance of GIC-D with $\rho=80$ is not degraded by the coherent signal mismatches.

Summarizing, the above analysis has singled out GIC-D with $\rho=80$ as the recommended detection architecture capable of providing a
satisfactory detection and classification performance in the presence of coherent signals at least for the considered scenarios.

\section{Conclusions}
\label{Sec:Conclusions}
In this paper, we focused on the adaptive radar detection in the presence of fully correlated signals besides that of interest.
Such additional signals may be due to multipath propagation effects or to the action of malicious platforms (smart jammers).
In order to account for different operating scenarios, at the design stage, we have considered a multiple-hypothesis test
that also includes the classical radar signal-plus-interference hypothesis and devised likelihood-ratio-based decision schemes
whose statistics under a specific hypothesis depend on a suitable penalty factor tuned according to the number of unknown parameters under that 
hypothesis (leveraging the approach of the MOS rules). As a result, such architectures are provided with classification capabilities returning,
as a byproduct, an estimate of the number of coherent signals impinging on the radar system. The performance analysis has been carried out resorting
to simulated data considering three different scenarios with an increasing number of coherent signals. Moreover, for comparison purposes,
the curves for the GAMF and GASD have been also reported. The analysis has singled out the GIC-D with $\rho=80$ as the recommended 
detection architecture since it overcomes the remaining proposed decision schemes in terms of both detection and classification performance.

Future research tracks may include the design of (possibly space-time) processing architectures that account for coherent 
signals spread along the range dimension or aimed at operating in scenarios where multiple coherent and/or uncorrelated signals are present.

\bibliographystyle{IEEEtran}
\bibliography{group_bib}
\end{document}